\documentclass{ar-1col-S2O}

\usepackage[comma]{natbib}
\usepackage{url}

\jname{Annu. Rev. Astron. Astrophys.}
\jvol{ARA\&A}
\jyear{2026}
\doi{10.1146/((please add article doi))}

\newcommand{\Msun}{\ensuremath{\,{\rm M}_\odot}}                  
\newcommand{\Rsun}{\ensuremath{\,{\rm R}_\odot}}                  
\newcommand{\Teff}{\ensuremath{T_{\rm eff}}}                      
\newcommand{\Porb}{\ensuremath{P_{\rm orb}}}                      

\usepackage{hyperref}  

\begin{document} 

\markboth{Southworth \& Bowman}{Pulsations in Binary Star Systems}                     

\title{Pulsations in \\ Binary Star Systems}                                  

\author{John Southworth$^1$ and Dominic M.\ Bowman$^{2,3}$
\affil{$^1$Astrophysics Group, Keele University, Staffordshire, ST5 5BG, UK; \\ email: taylorsouthworth@gmail.com}
\affil{$^2$School of Mathematics, Statistics and Physics, Newcastle University, NE1 7RU, Newcastle upon Tyne, UK; \\ email: dominic.bowman@newcastle.ac.uk}
\affil{$^3$Institute of Astronomy, KU Leuven, Celestijnenlaan 200D, B-3001, Leuven, Belgium}}

\begin{abstract}
High-precision and long-duration light curves from space telescopes have revolutionized the fields of asteroseismology and binary star systems. In particular, the number of pulsating systems in eclipsing binaries has drastically increased thanks to space-based observations covering almost the entire sky. When combined with multi-epoch spectroscopy, this allows us not only to measure model-independent dynamical masses and radii for thousands of eclipsing binary systems, but also facilitates the powerful synergy of binarity and asteroseismology. Moreover, asteroseismology of pre- and post-interaction binary stars allows the physics of binary evolution to be constrained, including tides, mass transfer, and even mergers.
\begin{itemize}
\item Eclipsing binaries are among the best laboratories for testing \\ stellar structure and evolution theory because we are able to \\ measure their masses and radii independently of models.
\item Combining binary and asteroseismic modeling yields precise \\ constraints on the physical processes at work within stellar \\ interiors, such as rotation and mixing.
\item Pulsating binaries are challenging to study given the plethora \\ of different techniques and physical processes that need to be \\ considered depending on their orbital and physical properties.
\item The impact of tides on the pulsational, stellar structure, and \\ orbital properties of a binary system can be tested through \\ tidal asteroseismology.
\end{itemize}
\end{abstract}

\maketitle

\tableofcontents


\section{INTRODUCTION}              
\label{section: intro}


Binary star systems consist of two gravitationally-bound stars orbiting each other. They provide crucial opportunities for understanding the physics of stars. The study of binary systems has a long history -- covered in many textbooks (e.g. \citealt{Hilditch01book}) -- and remains crucial in establishing how stars evolve (e.g.\ \citealt{DucheneKraus13araa, Offner+23aspc, Marchant++24araa}). The field has recently entered a new era due to high-quality data obtained by space missions. The same space photometry has also vastly improved our knowledge of stellar pulsations, which provide a direct method to constrain interior physical processes. The study of stellar pulsations -- asteroseismology -- delivers masses, radii and ages, but also constrains interior rotation, mixing, and magnetic fields (e.g.\ \citealt{Aerts21rvmp, Kurtz22araa}). When the techniques are combined, pulsations in binary systems provide the most precise and accurate observational constraints. In this review, we summarize the complementary advances in the study of binary stars and asteroseismology, and focus on pulsating binary systems as excellent laboratories for refining stellar evolution theory.

\subsection{History and definitions of the different types of binary stars}
\label{section: intro: binaries}

\begin{marginnote}[]
\entry{Visual binary}{Two stars with a small angular separation as seen by observers on Earth. The stars may or may not be gravitationally bound.}
\entry{Astrometric binary}{Two gravitationally bound stars with an orbit that can be spatially resolved by an observer on Earth.}
\end{marginnote}

\begin{marginnote}[]
\entry{Eclipsing binary (EB)}{Two stars which eclipse each other once or twice per orbit, causing a commensurate decrease in brightness of the system.}
\end{marginnote}

John \citet{Michell1767rspt} asserted that the distribution of stars on the celestial sphere was inconsistent with being random, and thus nearby stars were physically associated with each other. His arguments encompassed both \emph{visual binary} stars and the existence of clusters of stars such as the Pleiades. William Herschel presented a first catalog of double stars to the Royal Society in \citeyear{Herschel1782rspt}, and subsequently coined the term ``binary star'' \citep{Herschel1802rspt}. Herschel made the first detection of orbital motion outside our Solar system, for the Castor \emph{astrometric binary} system in \citeyear{Herschel1803rspt}. His observations of this system over 43\,yr suggested an orbital period of $\Porb = 342$\,yr, in reasonable agreement with modern measurements.

John \citet{Goodricke1783} discovered the periodicity of the changes in brightness of the star Algol ($\beta$\,Persei). He (very) briefly outlined two solutions: an \emph{eclipsing binary} (EB) system containing a large orbiting body, or inhomogeneities on the surface of a rotating star. Algol was already known to be variable: its decreases in brightness were mentioned in the Ancient Egyptian Cairo Calendar in approximately 1200 {\sc bce} \citep{JetsuPorceddu15plos}. \citet{Goodricke1785} subsequently found the second known EB, $\beta$\,Lyrae, and established its variability period but made no claim as to the cause. 

\begin{marginnote}[]
\entry{Spectroscopic binary (SB)}{A system where the spectral lines of two stars are observed.}
\entry{SB1 binary}{Spectral lines from one star are measurable with a periodic Doppler shift in wavelength caused by orbital motion.}
\entry{SB2 binary}{Spectral lines from both stars are measurable, each Doppler-shifted in wavelength with the same orbital period.}
\entry{Detached eclipsing binary (dEB)}{An eclipsing binary system where the stars are sufficiently far apart to have evolved as single stars.}
\end{marginnote}

A century later, \citet{Vogel1890pasp} proved Algol is a close binary system by showing that the radial velocity (RV) of the primary star was negative before, and positive after, primary eclipse. This work originated the term \emph{spectroscopic binary} (SB) to refer to objects that show spectral lines from two stars, which are likely Doppler shifted because of their orbital motion. At a similar time, Mizar ($\zeta$\,UMa) and Menkalinan ($\beta$\,Aur) were detected as \emph{double-lined spectroscopic binaries} (SB2) by Edward Pickering and Antonia Maury \citep{Pickering90mn}. \citet{Stebbins11apj} detected eclipses in $\beta$\,Aur, making it the first SB2 EB. His analysis yielded the first direct mass and radius measurements of stars other than our Sun, which were close to modern values \citep{Me++07aa}. This established that the masses and radii of stars can be measured directly if the system is an SB2 EB. The radii are obtained from the depth, duration and shape of the eclipses (see Section~\ref{section: analysis}), whereas masses are measured from the velocity semiamplitude of the stars' orbital motion via Newton's laws and Kepler's celestial mechanics. If one can deduce the effective temperature, $\Teff$, of each star, then their luminosities follow from $L=4\pi R^2\sigma_{\rm SB}{\Teff}^4$, where $R$ is radius and $\sigma_{\rm SB}$ is the Stefan-Boltzmann constant. If the stars in an EB are well separated, such that they have not experienced mass transfer or loss -- normally termed a \emph{detached eclipsing binary} (dEB) -- they are representative of single stars. Therefore, SB2 dEBs represent fiducial points against which to test the predictions of models of (single) stellar evolution.

\subsection{Photometric effects of binarity}
\label{section: intro: other}

The most distinctive indicator of binarity is \emph{eclipses}. The timing, depth, shape and duration of the eclipses depends on the radii of the stars, the orbital inclination and period, and eccentricity. The first rigorous analysis method for deducing the radii of stars from eclipse shapes was by Henry Norris \citet{Russell12apj} and there have been many advances since \citep[e.g.][]{Kopal59book}. Modern analysis methods allow measurement of these properties of a binary via numerical simulation of a system in a computer (see later).
Eclipses are more likely to occur in close binary stars, as a wider range of orbital inclinations result in eclipses visible to a distant observer. Other effects also become stronger in close binaries, can be detectable even in the absence of eclipses, and are called {proximity effects}:


The \emph{reflection effect} is the increase in brightness of a star's surface by irradiation from a close companion. This leads to an approximately sinusoidal brightness variation for each star in a binary throughout the orbital period; the effects from the two stars are anti-phased so cancel when the two stars are identical. The strength of the effect varies according to $(R/a)^2$, where $a$ is the semi-major axis of the relative orbit \citep{MorrisNaftilan93apj}. 

\begin{marginnote}
    \entry{Reflection effect}{Brightening of a star's surface due to irradiation from a companion.}
    \entry{Ellipsoidal effect}{Periodic brightening of a star due to gravitational distortion caused by a companion.}
    \entry{Doppler beaming effect}{Change in brightness of a star due its motion around a companion.}
    \entry{Apsidal motion}{Precession of an eccentric binary orbit.}
\end{marginnote}

The \emph{ellipsoidal effect} is a change in brightness of a star in a close binary where the gravity of its companion distorts it into an ovoid shape whose projected surface area varies during the orbit. This effect is approximately sinusoidal but at half the orbital period of the system; the effects from the two stars add together and are greatest when the stars are identical. The strength of the effect varies according to $(R/a)^3$ \citep{Morris85apj}.

\emph{Doppler beaming} is two effects which cause changes in the brightness of a star based on its RV. The first is that a star's orbital motion Doppler-shifts its spectrum, increasing or decreasing its brightness in the wavelength interval the observer is sensitive to. The second is the relativistic beaming of photons towards the direction of motion so a star appears brighter when approaching the observer. The two effects are additive and effectively imprint the RV curve of a star into a variation of its brightness. The Doppler beaming effect of the two stars are anti-phased so we observe only their difference (see e.g. \citealt{Bloemen+11mn}).


\emph{Apsidal motion} is the precession of the orientation of the orbit of an eccentric binary system, seen as a changing argument of periastron. This was first noticed in the Solar system in antiquity, and is caused by two mechanisms. There is a {classical} contribution due to the distortion of a body which is not a point mass \citep[e.g.][]{Russell28m}, and a {general-relativistic} contribution \citep{Levicivita37amjam}. The two add linearly, and can be quantified because they change the times of eclipses and the shape of the RV curve.

\subsection{Eclipsing binaries as tests of stellar theory}
\label{section: intro: theory}

Early stellar evolution theories concentrated on explaining the Hertzsprung--Russell (HR) diagram, which does not explicitly include mass information. The incorrect scenario originally outlined by \citet{Russell13obs} was that stars were born as giants, contracted to form B (i.e.\ \emph{early type}) stars, and decreased in brightness through the F, G, K, and M classes (i.e.\ \emph{late type}) as they aged. The masses, and in particular the densities, from a large number of EBs \citep{Shapley13apj} rendered this viewpoint untenable \citep{Russell14obs}; we now know the missing ingredient was thermonuclear fusion \citep{Bethe39phrv}. Since then, EBs have been used as fundamental checks on the predictions of stellar evolution theory.

\begin{marginnote}
    \entry{Early type}{Stars of spectral types O and B are traditionally referred to as early-type because of the incorrect evolutionary sequence proposed by \citet{Russell13obs}.}
    \entry{Late type}{Stars of spectral type F, G, K, and M.}
\end{marginnote}

Today, significant uncertainties remain within single-star evolution models for interior mixing of chemical species and angular momentum transport, even in the absence of tides or mass transfer. In particular, dEBs are useful laboratories for constraining dynamical processes that contribute to mixing and angular momentum transport, for example, waves, shear, and other types of instabilities (see \citealt{Aerts++19araa}). This is because dEBs contain two stars of known mass, radius and luminosity, but also have same age and initial chemical composition. Moreover, the model-independent masses and radii of dEBs significantly improve asteroseismic modeling to constrain interior properties, such as rotation and mixing (e.g.\ \citealt{Themessl+18mn, Johnston+19mn}).

Several other ingredients needed for theoretical stellar models can be constrained by binary systems, but are not discussed in detail in this review. These include helium abundances \citep{Metcalfe+96apj}, the helium-to-metal enrichment ratio $\Delta Y/\Delta Z$ \citep{Ribas+00mn,Valle+24aa}, and the radius discrepancy seen in low-mass stars \citep{Hoxie73aa,Torres13an}. For further information see the review paper by \citet{Torres++10aarv}.

\subsection{The discovery of plentiful eclipsing binaries from space photometry}
\label{section: intro: discovery}

Early discoveries of EBs were based on ground-based monitoring campaigns supplemented by photographic variability studies. More recent surveys have typically found EBs as byproducts of data obtained for astrometry (e.g.\ \emph{Hipparcos}; \citealt{Vanleeuwen+97aa}) and \emph{Gaia} \citep{Gaia16aa}, microlensing (e.g.\ OGLE; \citealt{Soszynski+16aca}), or the detection of transiting exoplanets with, for example, SuperWASP \citep{Norton+11aa}, HAT \citep{Latham+09apj}, and KELT \citep{Oelkers+18aj}. 

Huge advances have been made using the extensive archives of light curves from space missions in the last two decades \citep{Me21univ}. Data from the \textit{CoRoT} mission \citep{Auvergne+09aa} led to the detection of 2269 EBs from light curves of 177\,454 objects \citep{Deleuil+18aa}. A total of 2878 EBs were identified from \textit{Kepler} \citep{Kirk+16aj}, and 69,000 EBs and counting from the ongoing TESS mission \citep{Prsa+21apjs,IJspeert+21aa,IJspeert+24aa,Kostov+25apjs}. These activities have not only greatly increased the number of EBs known, but have vastly improved the quality of the light curves.

\begin{summary}[]
\noindent The detached EB catalog (DEBCat; \url{https://www.astro.keele.ac.uk/jkt/debcat/}; \citealt{Me15aspc}) includes all dEBs with mass and radius measurements to 2\% or better. At the time of writing (August 2025) it contains 363 systems.
\end{summary}

\subsection{A brief overview of stellar pulsations}
\label{section: intro: puls}

Essentially all stars have the necessary properties at some point in their evolution to self-excite waves within their interiors, which means that pulsations are commonplace across the HR~diagram (see \citealt{Kurtz22araa}). Stellar pulsations can be described using \emph{spherical harmonics}, such that a pulsation mode has a frequency, $\nu$, and a radial order, $n$, angular degree, $\ell$, and azimuthal order $m$. The goal of asteroseismology --- the study of stellar pulsations (e.g.\ \citealt{Aerts++10book}) --- is to identify the spherical harmonic geometry of observed pulsation frequencies, and use these as diagnostics to extract the properties of a star's interior (see review by \citealt{Aerts21rvmp}). Pulsation modes are standing waves and are extremely sensitive to a star's interior physical conditions, thus they constrain bulk properties such as mass and radius, but also dynamical processes such as mixing and rotation (see \citealt{Aerts++19araa}).

\begin{marginnote}[]
\entry{Spherical harmonics}{Pulsational geometry in spherical stars that is characterised by three quantum numbers: radial order $n$, angular degree $\ell$, azimuthal order $m$.}
\end{marginnote}

There are generally two main flavors of pulsation modes: (i) pressure modes (i.e.\ standing sound waves), which propagate in both radiative and convective regions with the pressure gradient acting as the restoring force; and (ii) gravity modes, which only propagate in radiative regions with buoyancy being the restoring force, such that gravity modes are evanescent in convective zones. For high-radial order pulsation modes (i.e.\ $|n| \gg \ell$), the asymptotic approximation enables mode identification of pulsation frequencies in terms of spherical harmonic geometry (see \citealt{UNNO_BOOK_2ed}). For example, pressure modes of the same angular order and azimuthal order but consecutive radial order have the asymptotic property of being equally spaced in a frequency spectrum, thus defining the \emph{large-frequency spacing}, $\Delta\nu$ (see \citealt{ChaplinMiglio13araa}). High-radial order gravity modes are instead asymptotically equally spaced in period leading to an \emph{asymptotic period spacing}, $\Delta\Pi$ (see \citealt{UNNO_BOOK_2ed}). It is therefore the structure of a star that defines the frequencies and propagation cavities of pulsation modes, with pressure and gravity modes being dependent on the local sound speed and buoyancy frequency, respectively \citep{Aerts++10book}. Rotation, magnetic fields, and binary companions mean that the Coriolis, Lorentz, and tidal forces, respectively, are also important to consider. This is because additional forces can modify a star's structure, thus perturb its pulsation frequencies \citep{Aerts21rvmp}.

\begin{marginnote}[]
\entry{Large frequency spacing ($\Delta\nu$)}{Inverse of the travel time of a sound wave along a star's diameter. It directly probes a star's average density.}
\entry{Asymptotic period spacing ($\Delta\Pi$)}{Inverse of the travel time of a gravity wave along a star's diameter. It probes a star's buoyancy, gravity and density.}
\end{marginnote}

Different types of pulsating stars are usually grouped based on having a similar mass, age, or type(s) of pulsation mode. For example, some groups of pulsators are named after the first star to show pulsations; for example $\beta$~Cep, $\delta$~Sct, and $\gamma$~Dor. However, other pulsator classes are more generally defined, such as stars with solar-like oscillations (SLOs). See \citet{Kurtz22araa} for a review.

\emph{Low-mass stars} have structures similar to the Sun, thus have birth masses smaller than about 1.2\Msun. Such stars have thick convective envelopes whilst on the main sequence, and their pressure-mode pulsations are stochastically excited by turbulent convection in their outer layers \citep{GarciaBallot2019}. This typically gives rise to pulsation mode frequencies spanning moderate-to-high radial orders, and allows $\Delta\nu$ and the frequency of maximum power, $\nu_{\rm max}$, to be identified, which are key diagnostics needed to perform global asteroseismology \citep{ChaplinMiglio13araa}. The later evolutionary stages of \emph{intermediate-mass stars} (i.e.\ birth masses between about 1.2 and 8\Msun), such as red giants, also have thick convective envelopes and exhibit stochastically-excited mixed pressure-gravity modes \citep{HekkerJCD2017aapr}. 

\begin{marginnote}[]
\entry{Low-mass stars}{Stars with birth masses below about 1.2\Msun, such that they are similar to the Sun.}
\entry{Intermediate-mass stars}{Stars with birth masses between about 1.2 and 8\Msun.}
\entry{Massive stars}{Stars with birth masses above 8\Msun, such that they are progenitors of neutron stars and black holes.}
\end{marginnote}

\emph{Massive stars} (i.e.\ birth masses above 8\Msun, hence progenitors of neutron stars and black holes) and intermediate-mass stars have radiative envelopes during the main sequence, and their pulsations are self-excited through a heat-engine mechanism operating in thin partial ionization zones \citep{MoskalikDziembowski92aa, Dziembowski++93mn, DziembowskiPamyatnykh93mn}. This opacity-driven excitation mechanism yields coherent pulsation modes with long lifetimes. However, the complexity of this mechanism in governing which pulsation modes are excited is not fully understood (see e.g.\ \citealt{Szewczuk++17mn}). Massive stars have a diverse range of pulsation mode geometries and variability timescales \citep{Bowman20faas}. They are on average faster rotators than low-mass stars \citep{Abt++2002apj, Royer++2007aa}, which impacts their structure and pulsations. Moreover, massive stars exhibit radiatively driven winds with large mass-loss rates (see review by \citealt{Puls++08}), and are also commonly found in multiple systems (e.g.\ \citealt{MoeDiStefano17apjs}). These processes complicate the asteroseismic analysis of massive stars. For example, rotation breaks the assumption of spherical symmetry making mode identification more challenging. On the other hand, binarity can be used as an advantage; for example, model-independent masses and radii of EBs as constraints in forward asteroseismic modeling (e.g.\ \citealt{Guo++17apj2, Johnston+19mn, Miszuda+21mn}).

\subsection{Historical discoveries of pulsating binary systems}
\label{section: intro: history}

The discovery of pulsations in EBs required ground-based photometric campaigns, but were rather serendipitous with any additional periodicity to the orbital period being identified as pulsations. To our knowledge the first reported pulsating EB was AB\,Cas, for which \citet{Tempesti71ibvs} found brightness fluctuations of period 1.4\,hr and amplitude 0.05\,mag. This was attributed to $\delta$~Sct pulsations (see section \ref{section: results: DSct}) in the A3 primary star. A recent detailed analysis has confirmed this picture, with the use of TESS space photometry data \citep{Miszuda+22mn}. Similarly, $\delta$~Sct pulsations with a dominant period of 1.5\,hr were detcted in the A8 primary of Y\,Cam \citep{BrogliaMarin74aa}. A detailed analysis of the system was performed using ground-based photometric data from multiple continents \citep{Rodriguez+10mn} and most recently space-based photometry \citep{CelikKahraman24pasj}. Both AB\,Cas and Y\,Cam are \emph{semi-detached} (Algol) systems in which the A-type primary star is a $\delta$~Sct pulsator. Not long after the discovery of these pulsating Algol systems, the first dEBs containing a $\delta$~Sct pulsator were found: AI\,Hya \citep{JorgensenGronbech78aa} then RS\,Cha \citep{ClausenNordstrom80aa}. 

\begin{marginnote}[]
\entry{First known pulsating EBs}{AB\,Cas is an Algol containing a $\delta$~Sct star, and V539\,Ara is a dEB containing an SPB star. Both were announced in 1971.}
\entry{Semi-detached}{A binary system where one star is filling its Roche lobe and the other is not.}
\entry{Algol system}{An eclipsing binary system where the secondary star fills its Roche lobe. The prototype is the star $\beta$\,Persei (Algol).}
\end{marginnote}

Although proximate to $\delta$~Sct stars in the HR~diagram, the first $\gamma$~Dor star in a binary system came much later, with the first being VZ\,CVn \citep{Ibanoglu+07mn}. Among B dwarfs, variability caused by pulsations in V539\,Ara was suggested by \citet{Knipe71aa} and confirmed by \citet{Clausen96aa}, making it the first known EB containing an SPB star. 
The number of EBs including SPB stars remains quite small: there are only a handful of confirmed cases (e.g.\ \citealt{MeBowman22mn}). This is partly a selection effect because the pulsation and orbital periods can be similar, making it hard to disentangle the brightness changes.

For massive stars, early discoveries of $\beta$~Cep pulsators in EBs include 16\,Lac \citep[also called EN\,Lac;][]{Jerzykiewicz+78ibvs,Jerzykiewicz+15mn}, $\lambda$~Sco \citep{Uytterhoeven+05aa}, and V381\,Car \citep{Freyhammer+05aa}. The first for which masses were precisely measured were V453\,Cyg \citep{Me+20mn} and VV\,Ori \citep{Me++21mn,Budding+24mn}. Probing binarity and pulsations for massive stars is challenging for two reasons: (i) massive stars commonly show stochastic low-frequency (SLF) variability in their light curves with periods similar to potential orbital periods \citep{Bowman+19natas}; and (ii) massive stars are intrinsically rare in the Universe. On the other hand, massive stars are luminous and their multiplicity fraction is high. The first well-studied example of an EB with SLF variability is V380\,Cyg \citep{Tkachenko+12mn,Tkachenko+14mn}, and a further eight examples were studied by \citet{MeBowman22mn} and \citet{Me23obs6}.  

The first ensemble analyses of pulsating binaries had to await modern space photometry missions. \citet{Debosscher+11aa} performed a variability survey of initial \textit{Kepler} mission data and found five EBs with additional variability typical of SPB or $\gamma$~Dor stars -- they did not distinguish between the two types due to the similar pulsation periods. Later, \citet{Debosscher+13aa} presented a detailed characterization of one of these pulsating EBs, KIC 11285625, comprising measurements of masses and radii of the component stars to 1\% precision and detection of several hundred pulsation frequencies. Many more are now known from space-based data \citep[e.g.][]{Li+20mnras2, MeVanreeth22mn}.

The \textit{Kepler} mission in particular revealed that red giant stars pulsate almost ubiquitously \citep[e.g.][]{DeRidder+2009nat, Hon+21apj}. Moreover, pulsating red giants in binary systems were found to be plentiful, with the first such system being KIC 8410637 \citep{Hekker+10apj}. Subsequent observations showed it to have a large orbital period (408.3\,d) and high eccentricity ($e=0.686$; \citealt{Frandsen+13aa}). Today, a large number of pulsating red giant binaries are known. For example, \citet{Gaulme+20aa} found evidence of binarity in 370 of 4500 red giants observed by the \textit{Kepler} mission. With recent \textit{Gaia} parallaxes and long-duration spectroscopic monitoring, the number of known long-period eccentric pulsating red giant binaries has grown dramatically \citep{Beck+24aa}.

Several more catalogs of pulsating stars in EBs have been published using data from space missions. These include 303 systems from \textit{Kepler} data \citep{GaulmeGuzik19aa} and several hundred from TESS \citep{Shi+22apjs,Chen+22apjs,Kahraman+22raa,Kahraman+23mn}. Catalogs of massive pulsators have been presented by \citet{MeBowman22mn} and \citet{EzeHandler24apjs} based on TESS data.

\section{ANALYSIS METHODS OF BINARITY AND PULSATIONS}
\label{section: analysis}

We are currently in the golden era of space mission surveys that provide high-precision, long-duration and rapid-cadence light curves for stars across the sky. Early space-based telescopes dedicated to studying stellar variability included the MOST \citep{Walker+03pasp}, BRITE-Constellation \citep{Weiss+14pasp}, and CoRoT \citep{Auvergne+09aa} missions, but these targeted relatively small fractions of the sky compared to the now-retired \textit{Kepler} \citep{Borucki+09sci} and ongoing TESS \citep{Ricker+15jatis} missions. \citet{Me21univ} provides a historical perspective of how space missions contributed to the discovery and analysis of binary systems. Here we describe the methods to characterize pulsating binary systems, and their photometric, spectroscopic, and asteroseismic analysis. We concentrate on EBs as these show the clearest indicator of binarity. However, eclipses depend on: (i) the orbital inclination of the system; and (ii) the sum of the radii of the stars expressed as a fraction of the orbital separation. This means that the likelihood of eclipses decreases for increasing \Porb, with only a minority of binary systems showing eclipses. 

\subsection{Photometric data reduction of eclipsing binaries}
\label{section: analysis: photometry}

\begin{marginnote}[]
\entry{Simple aperture photometry (SAP)}{Flux measurements of a star's brightness extracted from a target's pixel mask within each CCD image.}
\entry{Detrended light curve}{Time-series brightness measurements of a star that have undergone removal of instrumental systematics.}
\entry{Principal component analysis (PCA)}{Principal components are linear combinations of observables that maximally explain the variance of all variables.}
\end{marginnote}

Light curves are extracted from a series of CCD image data using \emph{simple aperture photometry} with a pixel mask that maximizes a target's flux whilst minimizing contamination from other stars. Light curves also need to be \emph{detrended} of instrumental systematics (e.g.\ \citealt{Jenkins+16spie}), which may include thermal changes of the instrument, and spatially and temporally varying sources of scattered light. A popular method of detrending is \emph{principal component analysis} \citep{Lightkurve+18})\footnote{\url{https://lightkurve.github.io/lightkurve/}}, but a more subjective approach of fitting a spline function to an extracted light curve is also feasible. Each approach has advantages and disadvantages and generally depends on the amplitude(s) and period(s) of the variability of a star. 

\begin{marginnote}
    \entry{Third light}{Flux contribution from stars other than in a binary system, for example background contaminating stars or a third star gravitationally bound to the binary system.}
\end{marginnote}

In the case of (pulsating) EBs, particular care is needed when the orbital period is similar to (or a fraction or multiple) of a light curve's duration; for example, a single TESS sector of 27.5\,d. In such cases, there is increased risk in artificially changing the shape of an eclipse via detrending, especially if performed automatically, or if only a single eclipse is present. If the pixel mask is not consistent (i.e. variable \emph{third light}), the depths of eclipses may be modified artificially which is difficult to manage in eclipse modeling. Finally, particular attention should be applied to pulsating EBs with shallow eclipses and/or high-amplitude pulsations, since the similar amplitude of these two types of variability signal can be challenging to unambiguously disentangle (see \citealt{MeBowman22mn} for examples).

\subsection{Detection of pulsating eclipsing binary systems}
\label{section: analysis: EBs}

\begin{figure} \centering
\includegraphics[width=0.995\textwidth]{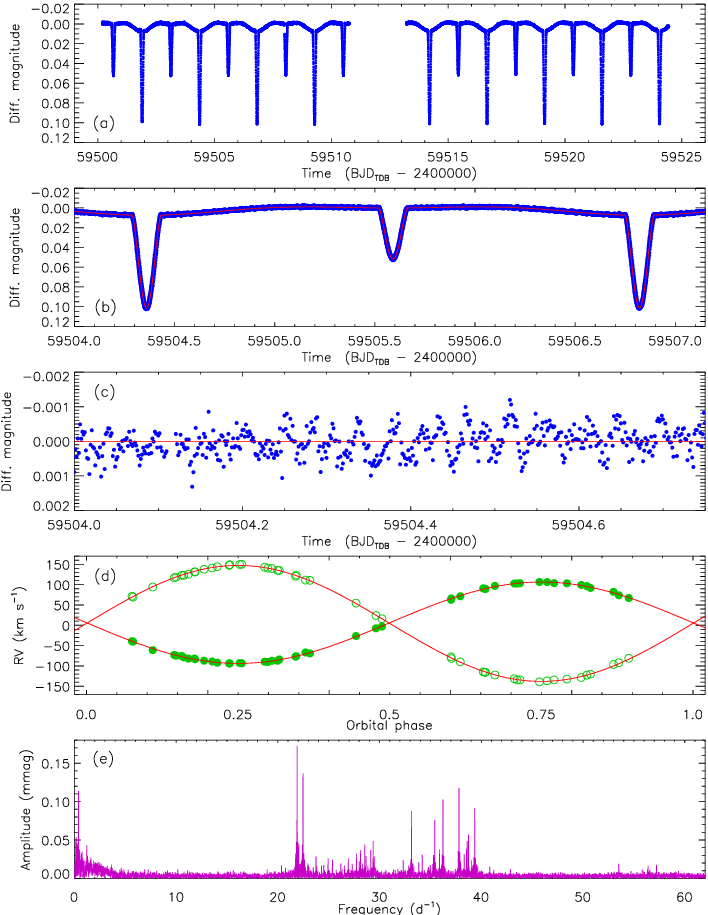}
\caption{Observations and analyses of the pulsating EB system HD\,23642 
\citep{Me++23mn}. Panel (a): light curve from sector 44 of TESS. Panel 
(b): close-up of part of the light curve (blue points) and the best fit 
to the data (red line) showing the pulsations. Panel (c): residuals of the 
fit to the light curve (blue points) around zero (red line). Panel (d): RVs 
of the primary (filled circles) and secondary (open circles) stars from 
\citet{Torres++21apj} compared to the best fits (red lines). Panel (e): 
amplitude spectrum of TESS sectors 43 to 45 after removal of the effects 
of binarity, showing multi-periodic $\delta$~Sct pulsations.}
\label{fig:23642}
\end{figure}

The discovery of pulsating EBs can be performed in a relatively automated way using eclipse detection routines \citep[e.g.][]{Prsa+11aj, Prsa+21apjs, Deleuil+18aa}, although the classification of the detected systems can be more difficult \citep[e.g.][]{Jayasinghe+18mn}. The eclipses can then be modeled and removed, or omitted entirely, from the data, and the Fourier transform reveals the residual pulsational variability. The detection of significant peaks in the Fourier transform within a frequency range typical for pulsations expected in one or both of the component stars thus means the system likely contains a pulsating star. An example of the TESS light curve and the residual light curve after the binary model has been removed to reveal the pulsations is shown in Fig.~\ref{fig:23642} for the pulsating EB HD\,23642 \citep{Me++23mn}. In this case, the $\delta$~Sct pulsations are at high frequencies of $20 < \nu < 60$~d$^{-1}$, thus far away in Fourier space from the orbital frequency (0.41~d$^{-1}$).

However, complications caused by imperfect detrending or astrophysical variability such as star spots can mimic eclipses or pulsations. In such cases, the Fourier transform may contain peaks at harmonics of the orbital frequency. For example, the orbital frequency arises from adjacent primary minima, but twice this value corresponds to the time difference between successive primary and secondary minima in a system with a circular orbit. Higher-order harmonics typically decrease in power for increasing frequency but are necessary due to the non-sinusoidal shape of the eclipses in the light curve. The orbital harmonics typically overlap with the gravity-mode frequency regime of SPB and $\gamma$~Dor pulsations. Therefore, careful attention is needed to isolate these signals; pulsations commensurate with the orbital frequency are more difficult to disentangle so automated searches for them are less efficient. For descriptions of these issues see \citet{IJspeert+21aa}.

\subsection{Modeling the photometric signatures of binarity}
\label{section: analysis: binary modeling}

Binarity most clearly manifests itself in a light curve as eclipses. Shorter-period systems may also show the ellipsoidal, reflection and/or Doppler beaming effects (see Section~\ref{section: intro: other}). There are two main motivations to model the photometric effects of binarity. First, this yields crucial information on the system, such as the orbital period and radii of the stars. Second, the effects of binarity can be removed from a light curve, leaving behind the pulsations and allowing asteroseismology \citep{Aerts++10book}.

\begin{marginnote}
   \entry{Roche potential}{\ \ The combined gravitational and rotational potential experienced by a test particle in a system containing two point masses in a circular orbit.}
   \entry{Orbital eccentricity}{A measure of how non-circular an orbit is. The limiting values for bound orbits are $e=0$ (circular orbit) and $e=1$ (radial orbit).}
   \entry{Argument of periastron}{The angle in the orbital plane between the ascending node and the periastron passage.}
\end{marginnote}

Binary modeling codes include {\sc WD} \citep{WilsonDevinney71apj}, 
{\sc jktebop} \citep{Me++04mn2}, {\sc phoebe2} \citep{Prsa+16apjs, Prsa18book}, and {\sc ellc} \citep{Maxted16aa}. The {\sc WD} code was the first widely used of these codes. It represents stellar surfaces as a mesh at a constant potential, specifically the \emph{Roche potential} modified to include rotation. This accounts for deformation of the stars due to their proximity (i.e.\ the ellipsoidal effect) and their mutual irradiation (i.e.\ the reflection effect). The main parameters to be fitted to the data are: the potentials of the two stars, which translate into their fractional radii ($r_{\rm A}$ and $r_{\rm B}$); \Teff\ values; orbital inclination ($i$); mass ratio ($q$), \emph{orbital eccentricity} ($e$); \emph{argument of periastron} ($\omega$); orbital period; and time of mid-eclipse ($T_0$). When stars are deformed it is standard practice to quote the volume-equivalent \emph{fractional radius}. The {\sc WD} code has a multitude of options which we do not discuss in this review. However, owing to its complexity, it can be quite slow and is limited by its numerical precision. Moreover, eccentric systems require significantly more time because the shapes of the stars must be calculated for every data point rather than once for all data points.

\begin{marginnote}
   \entry{Fractional radius}{The radius of a sphere of the same volume as a distorted star in a binary.}
\end{marginnote}

The currently most sophisticated code for modeling the effects of binarity is {\sc phoebe}\footnote{\url{http://phoebe-project.org/}} \citep{PrsaZwitter05apj}. This was originally based on the {\sc WD} code but has since been rewritten from scratch ({\sc phoebe2}; \citealt{Prsa+16apjs}) with extensive improvements to the physics implemented \citep{Prsa18book}.

Some other codes trade sophistication for speed. One such example is {\sc jktebop} \citep{Me13aa}, which is based on the {\sc ebop} code by \citet{Etzel75} and the spherical-star model of \citet{NelsonDavis72apj}. {\sc jktebop} is suitable when the stars are close to spherical (we recommend $r_{\rm A} < 0.2$ and $r_{\rm B} < 0.2$) and compensates for this restriction by being orders of magnitude faster than the {\sc WD} code. The ethos of {\sc jktebop} is to use parameters closely related to the shape of a light curve: $\Porb$, $T_0$, $r_{\rm A}$, $r_{\rm B}$, $i$, $e\cos\omega$ (which governs when the secondary eclipse happens), $e\sin\omega$ (which specifies the relative durations of the primary and secondary eclipses), and the ratio of the surface brightnesses (which gives the ratio of the eclipse depths). Several limb darkening laws are available, and the choice of which to use is unimportant in most cases \citep{Me23obs2}. The reflection and ellipsoidal effects are included using simple parameterizations, but Doppler beaming is not.

\subsection{Modeling the pulsations in the light curve}
\label{section: analysis: pulsations}

The traditional approach to modeling pulsating EBs has been to use a two-step methodology: (i) obtain a physical binary model; and (ii) subtract this binary model from the light curve, and analyze the residual light curve for pulsation mode frequencies using \emph{iterative pre-whitening} (e.g.\ \citealt{Bowman+19apj, Sekaran+20aa, Miszuda+22mn}). This process is illustrated for HD\,23642 in Fig.~\ref{fig:23642}, which is a pulsating EB containing a $\delta$~Sct star \citep{Me++23mn}. However, this approach generally works best when pulsation mode amplitudes are small relative to the eclipse depths, such that they can be treated as small perturbations (i.e.\ `noise') within the binary modeling. For larger pulsation amplitudes and/or smaller eclipse depths, this becomes increasingly less valid and binary modeling becomes more challenging. In extreme cases, it is difficult to construct a robust binary model and the two-step approach needs to be reversed: the pulsations are identified and removed from the light curve prior to binary modeling. Manually excluding the shallow primary eclipses allows frequency analysis of the high-amplitude pulsations to be performed, which can then be removed prior to binary modeling in an iterative fashion.

\begin{marginnote}[]
\entry{Iterative pre-whitening}{Removal of pulsations from a light curve based on a (co)sinusoid function with a frequency, amplitude and phase optimized by a least-squares fit to the light curve.}
\end{marginnote}

The ideal approach would be to physically model the impact of pulsations and eclipses simultaneously in a light curve as well as line profile variations (LPVs) caused by pulsations and RV variability caused by binarity. However, this would require a binary modeling code including (large) perturbations caused by pulsations described using spherical harmonics in the Roche geometry of the pulsating star. In the absence of a Roche geometry modeling framework that directly includes pulsations, an analytical approach to extracting the frequencies of pulsations and binary properties (e.g.\ \Porb\ and $e$) of pulsating EBs using Fourier analysis is a pragmatic compromise (e.g.\ \citealt{IJspeert+21aa, MeBowman22mn}). After pulsation frequencies have been extracted for a binary, the methods of asteroseismology can be used probe a star's interior physics. For dEBs it is reasonable to assume each component has evolved up to this point as a single star. However, for post-interaction binaries, more sophisticated forward asteroseismic modeling methods are needed that include the physics of mass transfer (e.g.\ \citealt{Miszuda+21mn, Miszuda+22mn}).

\subsection{Spectroscopy to determine fundamental parameters and pulsations}
\label{section: analysis: spectra}


\subsubsection{Masses from radial velocities}
\label{section: analysis: spectra: RVs}

To determine the masses of both component stars, a set of spectra with more epochs than the number of free parameters to be determined and which quasi-uniformly sample the orbital phase can be used to extract an RV time series. The RVs are then fit with a Keplerian orbit to determine the velocity semiamplitudes of the stars, $K_{\rm A}$ and $K_{\rm B}$. From $K_{\rm A}$ and $K_{\rm B}$, and the $\Porb$, $i$ and $e$ from the eclipse analysis, $a$ and the masses of the stars can be measured using standard formulae \citep[see][]{Hilditch01book}. From the eclipse modeling of the light curve, the $a$, $r_{\rm A}$ and $r_{\rm B}$ then give the true radii of the stars directly. An example of an RV time series and model fit for the pulsating EB HD\,23642 is shown in Fig.~\ref{fig:23642}.
When high-quality RV and light curve data are available, masses and radii can be measured to better than 1\% accuracy. These are empirical measurements as they depend almost entirely on geometry and celestial mechanics, and can be independent of models. For example, \citet{Maxted+20mn} achieved precisions of 0.1\% for the masses and radii of the dEB AI\,Phe, corroborated by multiple independent analyses of the data. 


\subsubsection{Atmospheric parameters}
\label{section: analysis: spectra: params}

Spectroscopic data allow one to constrain the $T_{\rm eff}$, surface gravity ($\log\,g$), and projected rotational velocity ($v\,\sin\,i$) of each component if its spectral lines are visible \citep{GRAY_BOOK}. This can be achieved through a statistical comparison of an observed spectrum, or a series of stacked spectra having been corrected for their individual RV offsets, to a grid of theoretical spectra calculated using a model atmosphere code. However, the spectroscopic analysis of SB2 binaries often requires techniques that separate the contribution of each component (e.g.\ spectral disentangling; \citealt{SimonSturm94aa}). The different numerical approaches to atmospheric modeling is beyond the scope of this review, but the relative precision for $T_{\rm eff}$ and $\log\,g$ can reach 1\% in the best cases. 

\subsubsection{Pulsations from spectral line profile variability}
\label{section: analysis: spectra: LPV}

Photometry is the most common method for studying pulsations, but time-series spectroscopy yielding LPVs is also an effective approach. Pulsations cause perturbations to spectral lines that can be extracted to create a time series, with a Fourier transform revealing any pulsation frequencies (see e.g.\ \citealt{Zima08comms}). However, the quality and number of spectra needed is immense, with high-resolution ($R > 50\,000$), high signal-to-noise (S/N $>$ 300), rapid cadence (of the order of minutes,) and long-duration time series spanning months typically needed for early-type stars (e.g.\ \citealt{Aerts+94aa, Uytterhoeven+04aa}). This can be challenging to assemble through competitive calls with modern observatories.

\subsection{The advantages of combining binarity and pulsation modeling}
\label{section: analysis: binary pulsations}

There is great synergy in combining the modeling of pulsations and binarity. Above all, the application of Kepler's laws to EBs yields precise and model-independent masses and radii that can be used as constraints in subsequent forward asteroseismic modeling studies. In this sense, forward asteroseismic modeling is complementary to binary modeling since it provides constraints on interior processes such as rotation, mixing, and angular momentum transport. Therefore, pulsating EBs are ideal laboratories that allow binary asteroseismology. Fortunately, the data demand for analyzing pulsations and EBs is the same: long duration, high-cadence, high-precision time-series photometry. 

A notable example of binary asteroseismology in action is KIC 10080943, which is a doubly-pulsating SB2 system with $M_1 = 2.0 \pm 0.1$\Msun, $M_2 = 1.9 \pm 0.1$\Msun, $R_1 = 2.9 \pm 0.1$\Rsun, and $R_2 = 2.1 \pm 0.2$\Rsun\ in a 15.34-d orbit with eccentricity of 0.449 \citep{Schmid+15aa}. Both stars show both gravity and pressure modes, which allowed their radial rotation profiles to be constrained \citep{SchmidAerts16aa}. 

In general, much of the asteroseismic literature has focused on single stars, with only a few, yet powerful, examples of forward asteroseismic modeling being applied to pulsating EBs. For dEBs, which can be effectively treated as single stars with a common age, dynamical masses and radii from binarity effectively lift degeneracies in forward asteroseismic modeling, for example, see \citet{Themessl+18mn}, \citet{Johnston+19mn}, and the aforementioned HD\,23642 \citep{Me++23mn}. There are significant challenges in combining asteroseismic and binary modeling techniques for interacting binary systems (see Section~\ref{section: impact: mergers}), but the scientific payoff is huge when successful.

\subsection{Which star is pulsating?}
\label{section: analysis: which}

\begin{figure} \centering
\includegraphics[width=0.98\textwidth]{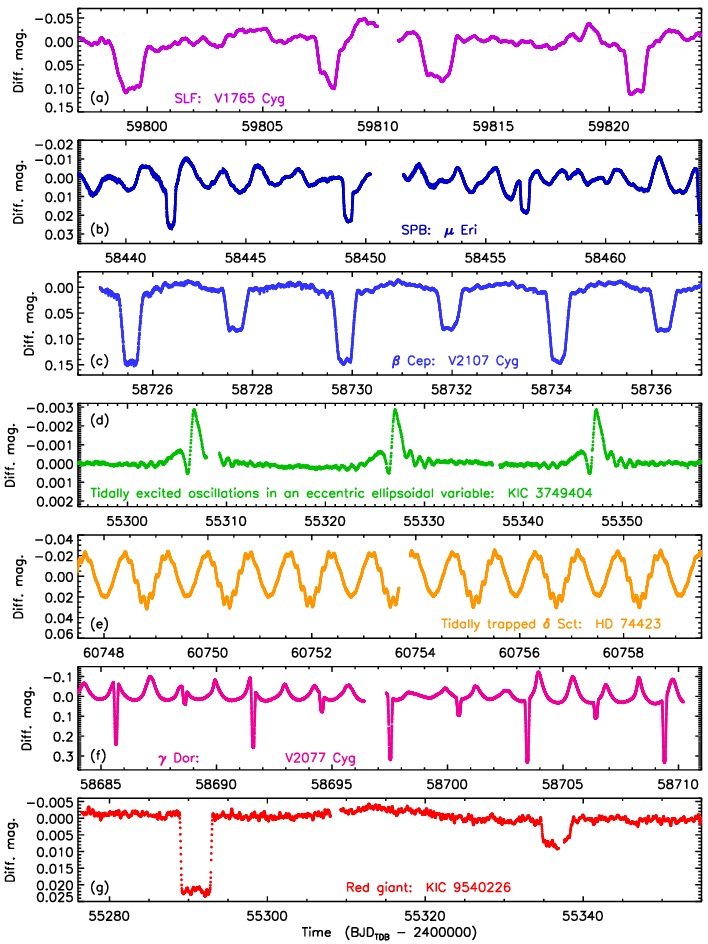}
\caption{Example light curves from the \textit{Kepler} and TESS space missions of EBs 
containing different pulsator classes. From top to bottom the panels show: SLF variability 
\citep{Me23obs6}, gravity-mode pulsations in an SPB star \citep{MeBowman22mn}, pressure-mode 
pulsations in a $\beta$~Cep star \citep{MeBowman22mn}, tidally excited oscillations (TEOs) 
in an eccentric ellipsoidal variable \citep{Thompson+12apj}, an ellipsoidal variable with 
tidally trapped $\delta$~Sct pulsations \citep{Handler+20natas}, gravity-mode pulsations 
in a $\gamma$~Dor star \citep{MeVanreeth22mn}, and a red giant \citep{Gaulme+14apj}.}
\label{fig: example LCs}
\end{figure}

\begin{figure}
\centering
\includegraphics[width=\textwidth]{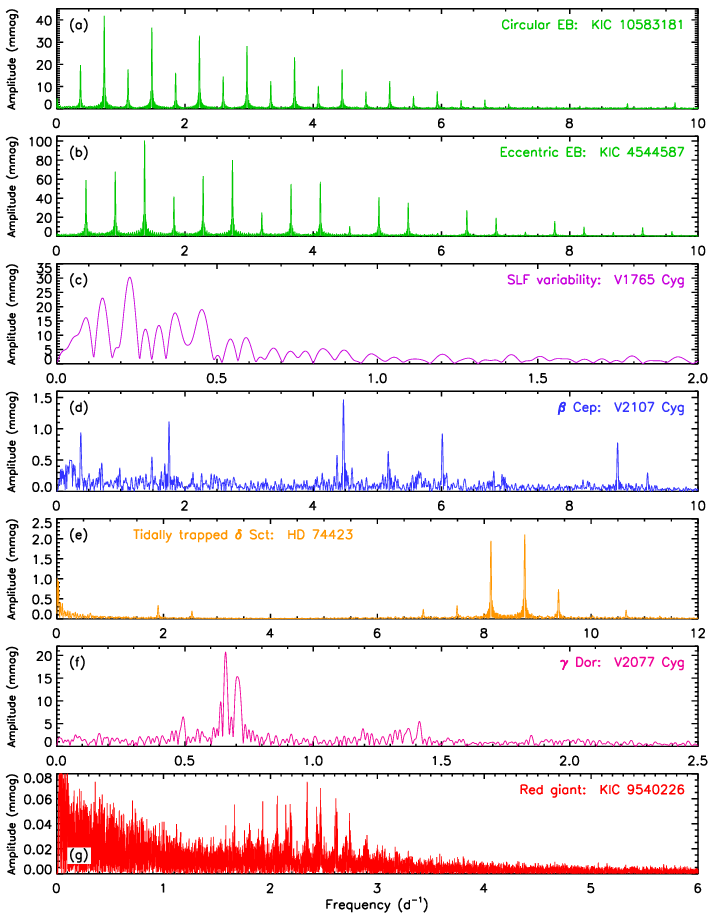}
\caption{Example frequency spectra for eclipses and pulsations. 
Panels (a) and (b) show frequency spectra of two EBs with deep eclipses: 
KIC 10583181 \citep{Helminiak+19mn} with a 2.70-d period circular orbit; 
and KIC 4544587 \citep{Hambleton+13mn} with a 2.19-d period eccentric orbit. 
Panels (c) to (g) show frequency spectra for a subset of the stars in 
Fig.~\ref{fig: example LCs} and were calculated after removal of the 
effects of binarity.}
\label{fig: example freq}
\end{figure}

Photometric observations of an unresolved binary star system contain the combined light of both stars. If pulsations are detected, it is not straightforward to determine which star is pulsating. The following methods, in no particular order, may be used to ascribe individual pulsation frequencies to a specific star in a binary system:
\begin{enumerate}
    \item The properties of the stars place one within and one well outside the appropriate instability region for pulsation in the HR~diagram. As an example, the SLOs in KIC 9540226 (Figs.\ \ref{fig: example LCs} and \ref{fig: example freq}) must originate from the red-giant component as those of its dwarf companion would be at much higher frequency. This method does not work for a binary system composed of two similar stars.
    \item One star may dominate the light of the system, and the amplitude of the pulsations exceeds the total light produced by the fainter star.
    \item The pulsation amplitudes change during eclipse. The amplitudes of pulsations appear weaker when it is eclipsed, and stronger when it is eclipsing its companion, due to the change in the fraction of light coming from that star (see Section~\ref{section: impact: tidally perturbed}).
    \item The pulsations show a light travel-time effect commensurate with the orbital motion of one of the component stars (see Section~\ref{section: results: pulsation timing}).
    \item Spectral LPVs are seen for one star and not for its companion.
    \item RV measurements show a contribution from pulsations for a specific star. A good example of this is GK~Dra \citep{GriffinBoffin03obs}, where the more massive component shows $\delta$~Sct pulsations which cause an approximately sinusoidal change in RV with an amplitude of 2.6\,km\,s$^{-1}$ in addition to the orbital RV motion.
\end{enumerate}


\section{PULSATING STARS IN BINARY SYSTEMS}
\label{section: results}

Here we review the main pulsator classes studied within EB systems, starting with the most massive stars and progressing towards lower mases. \citet{Lampens21galax} provides a non-exhaustive review of pulsators in EBs, and detailed discussions on asteroseismology and forward asteroseismic modeling of single stars are provided by \citet{Kurtz22araa} and \citet{Aerts21rvmp}, respectively. 

\subsection{Stochastic low-frequency variability in massive stars}
\label{section: results: SLF}

Space photometry has revealed that essentially all dwarf, giant, and supergiant massive stars exhibit SLF variability in their light curves \citep{Bowman+19natas}, which manifests across a broad period range from several days to minutes with variable photometric amplitudes \citep{Bowman+19natas}. SLF variability probes the mass and age of a massive star and is present across a wide range in metallicity \citep{Bowman+20aa, Bowman+24aa}. The physical mechanism for SLF variability is still debated, but stochastic gravity waves (i.e.\ damped gravity modes with finite lifetimes) excited by convection is likely based on comparison of observations and hydrodynamical simulations \citep{Bowman23apss}.

Studying SLF variability in binary stars is challenging since a light curve contains flux from both stars with potentially an unknown ratio, thus impacting its characterization. SLF variability has only been characterized in a handful of massive binary systems. The first example was V380\,Cyg, which contains a $11.4 \pm 0.2$\Msun\ primary and $7.0 \pm 0.1$\Msun\ secondary \citep{Tkachenko+14mn}. The time scales and amplitudes of the SLF variability for V380\,Cyg were similar to those required to explain spectroscopic LPVs caused by gravity modes \citep{Tkachenko+12mn}. Another notable example is V1765\,Cyg (Figs.\ \ref{fig: example LCs} and \ref{fig: example freq}), which contains a B0.5\,Ib primary and B1\,V secondary with dynamical masses of $22 \pm 2$ and $12 \pm 1$\Msun, respectively, in a 13.4-d orbit \citep{Me23obs6}. V380\,Cyg and V1765\,Cyg are among the few EBs with SLF variability and known masses and radii.

\subsection{$\beta$~Cep stars}
\label{section: results: BCep}

The $\beta$~Cep stars are pulsating massive stars with spectral types ranging from late-O to early-B on the main sequence, therefore they have birth masses typically between 8 and 25\Msun, but can be upwards of 30\Msun\ \citep{StankovHandler05apjs, Bowman20faas}. The heat-engine mechanism operating in the iron-bump opacity region is efficient at exciting low-radial order pressure and gravity modes with periods of order several hours \citep{MoskalikDziembowski92aa}. Modern space photometry, such as from the TESS mission, has demonstrated that $\beta$~Cep pulsations are relatively common among massive stars and they are found across a wide range of masses and ages \citep{Burssens+20aa, Fritzewski+25aa}. 

The first $\beta$~Cep star discovered in an EB was 16~Lac \citep{Jerzykiewicz+78ibvs}, a near-circular dEB with $\Porb = 12.1$\,d and high-amplitude pulsations \citep{DziembowskiJerzykiewicz96aa, Aerts+03aa, Jerzykiewicz+15mn}. Another famous example is $\lambda$~Sco, which is a hierarchical triple system comprising a pair of B stars with $\Porb = 5.95$\,d and $e=0.26$, and a tertiary star in an 1082-d orbit with eccentricity $e_3 = 0.23$ \citep{Uytterhoeven+04aa}. The first two $\beta$~Cep pulsators in EBs with precisely measured masses were V453\,Cyg and VV\,Ori \citep{Me+20mn,Me++21mn}. A sample of the TESS light curve of V2107\,Cyg in which deep primary and secondary eclipses and out-of-eclipse variability caused by pulsations can be seen in Fig.~\ref{fig: example LCs}. Analysis of both $\beta$~Cep primaries in V453\,Cyg and VV\,Ori revealed that tides in these short-period EB systems are strong enough to impact the pulsation mode frequencies (see Section~\ref{section: impact: tidally perturbed}). Moreover, VV\,Ori has significant evidence for gravitational interactions between the inner binary and a tertiary component: its light curve shows a changing inclination and different eclipse shapes over 50 years.

Recently, \citet{EzeHandler24apjs} published a catalog of 78 EBs containing a $\beta$~Cep pulsator, 59 of which are new discoveries, using TESS data of 8055 stars with spectral types between B0 and B3 for all luminosity classes. This represents a huge increase in the number of known $\beta$~Cep stars in EBs. The \Porb\ values range from 1.5\,d to over 30\,d, but the number of systems at longer \Porb\ falls off dramatically as most stars were only observed continuously for a single 27.5-d TESS sector and only a small range or orbital inclinations result in eclipses.


\subsection{Slowly pulsating B stars}
\label{section: results: SPB}

SPB stars are main-sequence stars between B3 and B9, so have birth masses between about 3 and 8\Msun. SPB stars have low-angular degree, high-radial order gravity modes with periods spanning a few days to several hours excited by the heat-engine mechanism \citep{Dziembowski++93mn}. Asteroseismology of a few dozen SPB stars has revealed their interior rotation rates, ranging from near-zero to near-critical, and a diverse range in their interior chemical mixing efficiencies \citep{Pedersen+21natas}.

There are very few SPB stars confirmed to be members of EB systems. Two (candidate) SPBs in EBs are V539\,Ara with a B3\,V primary and a B4\,V secondary in a 3.169-d orbit and $\mu$\,Eri with a B5\,IV primary and unknown secondary in a 7.4-d orbit \citep{Jerzykiewicz+13mn, MeBowman22mn}. A sample of the TESS light curve of $\mu$\,Eri, with primary eclipses and strong out-of-eclipse variability, is shown in Fig.~\ref{fig: example LCs}. The reason for the classification of candidate SPB stars for such systems is based on the limited number of TESS sectors providing insufficient frequency resolution for the close-frequency spacing of gravity modes. Moreover, the potentially similar pulsation, rotation, and orbital periods of such systems make it difficult to distinguish different types of variability. Automated searches of space mission light curves are effective at finding gravity-mode pulsators (e.g.\ \citealt{Debosscher+11aa, IJspeert+21aa}), but are usually unable to distinguish SPB and $\gamma$~Dor pulsators without additional information. 



\subsection{$\delta$~Sct stars}
\label{section: results: DSct}

The $\delta$~Sct pulsators are found at the intersection of the classical instability strip and the main sequence in the HR~diagram, range in spectral type from A2 to F5, and consist of both dwarfs and subgiants \citep{Breger00aspc, RodriguezBreger01aa}. This means $\delta$~Sct stars have birth masses between approximately 1.5 and 3\Msun\ \citep{BowmanKurtz18mn, Murphy+19mn}. The heat-engine mechanism operating in the He\,{\sc ii} ionization zone is efficient at exciting low-radial order pressure modes in $\delta$~Sct stars \citep{Chevalier71aa}, but they commonly also exhibit gravity modes that are difficult to explain theoretically without time-dependent convection models \citep{Dupret+05aa}. The 2-min cadence of the TESS mission revealed a subset of young $\delta$~Sct stars with high-frequency pulsations with $\Delta\nu$ spacings \citep{Bedding+20nat}, which may be excited by turbulent pressure since this has been shown to be necessary to excite high-frequency high-radial order pressure modes \citep{Antoci+19mn}.

Since $\delta$~Sct stars are common among A and F stars, they have a rich history and many have been discovered to be in binary systems (e.g.\ \citealt{LiakosNiarchos17mn}). Fig.~\ref{fig:23642} shows the example of HD\,23642, a member of the Pleiades open cluster, whose TESS light curve shows shallow eclipses, a reflection effect, and $\delta$~Sct pulsations with frequencies between $20 < \nu < 60$\,d$^{-1}$. Modeling of the eclipses and RVs yielded masses of $M_1 = 2.27 \pm 0.01$\Msun\ and $M_2 = 1.60 \pm 0.01$\Msun, and radii of $R_1 = 1.80 \pm 0.02$\Rsun\ and $R_2 = 1.41 \pm 0.02$\Rsun \citep{Me++23mn}. Asteroseismic modeling of the $\delta$~Sct pulsations yielded an age of $170 \pm 20$\,Myr and an envelope rotation period of 2.46\,d, which is consistent with the spectroscopic \Porb\ and $v\,\sin\,i$ constraints. For post-interaction binaries containing a $\delta$~Sct pulsator, such as KIC 10661783 and AB\,Cas, asteroseismology can constrain the efficiency of mass transfer and angular momentum transport \citep{Me+11mn,Miszuda+21mn, Miszuda+22mn}. 

\begin{marginnote}[]
\entry{L1 Lagrange point}{The location in space between two gravitationally bound objects at which there is a balance of the gravitational forces of both bodies and centrifugal force due to the orbital motion.}
\entry{Blue straggler}{A star that appears bluer and/or has a higher mass than the main-sequence turn-off of a cluster.}
\end{marginnote}

\subsubsection{Oscillating Eclipsing Algol (oEA) stars}

\citet{Mkrtichian+02aspc,Mkrtichian+03aspc} defined a subclass of $\delta$~Sct stars in mass-accreting Algol-type EBs called oscillating eclipsing Algol (oEA) stars. Such systems are thought to have undergone (or are undergoing) mass transfer through Roche lobe overflow at the \emph{first Lagrangian point} (L1), with the mass accreted by the $\delta$~Sct star having rejuvenated it to become a \emph{blue straggler}\footnote{The term field blue straggler can be used for blue stragglers not in clusters, such as oEA stars.}. There is an empirical correlation between the orbital periods and the dominant pulsation periods of $\delta$~Sct stars in binary systems with orbital periods below about 13\,d \citep{LiakosNiarchos17mn, Kahraman+17mn}. For short-period Algols, the pulsations may shed light on how recent mass transfer took place, because the pressure modes of $\delta$~Sct stars probe average density (see review by \citealt{Guo21frontiers}). The pressure modes of young $\delta$~Sct stars can reach frequencies as high as 80\,d$^{-1}$, which decrease in frequency as a star ages. For example, evolved $\delta$~Sct stars typically have low frequencies (i.e.\ $5 < \nu < 15$\,d$^{-1}$). The higher pulsation frequencies found in oEA systems support this rejuvenation hypothesis (see Section~\ref{section: impact: mergers}). 

\begin{marginnote}[]
\entry{Chemically peculiar stars}{Stars whose spectra show strange chemical abundance patterns due to atomic diffusion or magnetic fields.}
\end{marginnote}

\subsubsection{Chemically peculiar $\delta$~Sct stars}

The chemically peculiar Ap stars are rarely found in (close) binaries despite occupying a similar parameter space as the $\delta$~Sct stars in the HR~diagram \citep{Kurtz22araa}. However, \citet{Skarka+19mn} reported HD\,99458 to be the first Ap star with $\delta$~Sct pulsations in an EB, having $\Porb = 2.722$\,d and grazing eclipses. This is noteworthy for two reasons: (i) it is rare to find an Ap star in a short-period binary (see e.g.\ \citealt{AbtSnowden73apjs}), let alone a pulsating Ap star; and (ii) the light curve also showed signatures of spots hinting at the presence of a large-scale magnetic field, which is also rare in $\delta$~Sct stars \citep{Thomson-Paressant+23mn}.

\subsubsection{Non-eclipsing $\delta$~Sct binaries from pulsation timing}
\label{section: results: pulsation timing}

The light travel-time effect of a pulsating binary system is an established technique for studying non-eclipsing systems (see \citealt{Sterken05aspc}). \citet{ShibahashiKurtz12mn} and \citet{Shibahashi++15mnras} showed how the frequencies of pulsation modes are Doppler shifted, thus yielding a near-complete description of the orbit. An analogous phase modulation technique of pulsation modes has been applied to thousands of $\delta$~Sct stars observed by the \textit{Kepler} mission, yielding a few hundred non-eclipsing binary $\delta$~Sct stars \citep{Murphy+14mn, Murphy++16mnras, Murphy+18mn}. Recent methodological improvements include forward-modeling the light travel-time effect in the light curve \citep{Hey+20aj}, as opposed to dividing a light curve and measuring the phase modulation of individual pulsation modes. 

A notable system that combines pulsation timing and asteroseismology is KIC 9773821, which is an A dwarf and red giant binary with $\Porb = 481.9$~d and $e = 0.241$ \citep{Murphy+21mnras_2}. The \textit{Kepler} light curve revealed $\delta$~Sct pulsations with a measurable light travel-time  effect, as well as a pulsating red giant companion. The combination of binary modeling and forward asteroseismic modeling yielded a precise age of $1.08^{+0.06}_{-0.24}$~Gyr, as well as each star's evolutionary history \citep{Murphy+21mnras_2}.

\subsection{$\gamma$~Dor stars}
\label{section: results: GDor}

The $\gamma$~Dor stars comprise late-A to early-F dwarfs \citep{Kaye+99pasp,Tkachenko+13aa}, which pulsate in high-radial-order gravity modes excited through a heat-engine-like flux-modulation mechanism operating in their convective envelopes \citep{Guzik+00apjl, Dupret+04aa}. \textit{Kepler} mission data revolutionized asteroseismology of $\gamma$~Dor stars and has provided accurate masses and ages \citep{Mombarg+19mn}. Additionally, asteroseismology demonstrated quasi-rigid interior rotation profiles for hundreds of $\gamma$~Dor stars \citep{Vanreeth+16aa, VanReeth+18aa, Li+20mnras}, which current angular momentum theory struggles to explain \citep{Ouazzani+19aa, Aerts++19araa}. 

\citet{Debosscher+11aa} identified candidate SPB and $\gamma$~Dor pulsators in EBs using automated eclipse-finding algorithms applied to \textit{Kepler} light curves. But, as mentioned previously, it is difficult to ascertain if a pulsator is an SPB or $\gamma$~Dor pulsator without additional constraints, such as color or $T_{\rm eff}$ from spectroscopy. \citet{Sekaran+20aa} and \citet{Li+20mnras2} have published catalogs of about 100 $\gamma$~Dor pulsators in EBs using {\it Kepler} data, showing they are relatively common. A recent example of a well-characterized $\gamma$~Dor in an EB with deep eclipses is V456\,Cyg \citep{Vanreeth+22aa}, which contains stars of mass 1.9 and 1.6\Msun. High-amplitude gravity-mode pulsations in close binaries can make binary modeling challenging because of the similarity between pulsation and orbital periods, and between pulsation amplitudes and eclipse depths, e.g.\ V2077~Cyg (Fig.~\ref{fig: example LCs}).





\subsection{Solar-like oscillations in dwarfs}
\label{section: results: SLO dwarfs}

The lowest-mass dwarf star with detected SLOs is $\epsilon$\,Indi\,A, which is also in a high-order multiple system with $\epsilon$\,Indi\,Ba and Bb being brown dwarfs, and $\epsilon$\,Indi\,Ab being a Jupiter-like exoplanet \citep{Lundkvist+24apj, Campante+24aa}. The measured frequency of maximum power for $\epsilon$\,Indi\,A is $\nu_{\rm max} = 5265$\,$\mu$Hz (the highest detected for SLOs), and its large frequency spacing is $\Delta\nu = 201.25$\,$\mu$Hz, from which a mass of 0.78\Msun\ and an upper limit on age of 4\,Gyr was found.

Another notable example of SLOs in a dwarf star in a binary system is 16\,Cyg A and B. The pulsations in this pair of solar-analogs have been studied extensively (e.g.\ \citealt{SilvaAguirre+17apj}), especially since the release of the \textit{Kepler} mission light curve, making it a member of the \textit{Kepler} legacy sample of dwarf stars with SLOs \citep{Lund+17apj}. Structure inversions of the pulsation frequencies have allowed precise estimates of the radii and masses of 16\,Cyg\,A and B, but reveal some differences in the sound speed profiles when compared to the best-fitting results from forward modeling based on grids of evolutionary models \citep{Bellinger+17apj, Buldgen+22aa}.

\subsection{Solar-like oscillations in giants}
\label{section: results: SLO giants}

The \textit{Kepler} mission revolutionized asteroseismology of giant stars with SLOs, with tens of thousands of pulsating red giants being discovered \citep{ChaplinMiglio13araa}. KIC 8410637 was the first pulsating red giant in a EB to be discovered \citep{Hekker+10apj} and characterized \citep{Frandsen+13aa}. Since then, about 1000 pulsating red giants in binaries have been studied \citep{Gaulme+13apj,Gaulme+14apj,Themessl+18mn,Beck+14aa,Beck+24aa}, which span the same mass range (i.e. $0.5 < M < 3.5$~M$_{\odot}$) as the complete \textit{Kepler} red giant sample. An extract of the 4-yr \textit{Kepler} light curve of KIC 9540226, a red giant in an EB with $\Porb = 175.46$\,d and $e = 0.39$ \citep{Gaulme+14apj}, is shown in Fig.~\ref{fig: example LCs}.

Pulsating red giants in binaries have orbital periods ranging from days to over a decade and include near-circular to highly eccentric systems. They offer the chance to check if the masses and radii of red giants estimated from asteroseismology agree with dynamical values. Agreement between these methods has been found by some authors after applying corrections to the asteroseismic values \citep{Themessl+18mn,Brogaard+18mn}, whereas others have shown that the asteroseismic masses and radii are too large by 15\% and 5\% respectively \citep{Gaulme+16apj,Benbakoura+21aa}. Recent works by \citet{Brogaard+22aa} and \citet{Thomsen+25aa} have found good agreement between asteroseismic and dynamical properties for the metal-poor systems KIC 4054905 and KIC 10001167.


Tides are observed to suppress pulsation mode amplitudes in some red giant binaries: \citet{Gaulme+16apj} found that pulsations were not detected in systems whose radii summed to more than 16\% of the orbital separation. \citet{Beck+24aa} showed that the \emph{equilibrium tide} impacts the periods and eccentricities of red giant binaries (see also Section~\ref{section: impact: tidally perturbed}). 

\begin{marginnote}
    \entry{Equilibrium tide}{the balanced tidal force exerted between a pair of stars in a circular and synchronized binary, leading to time-independent deformation.}
\end{marginnote}

\citet{Deheuvels+22aa} demonstrated how asteroseismology can reveal if a red giant has interacted with a companion. Specifically, the properties of the cores of red giants, and whether they are degenerate or not, can be probed through their location in the $\Delta\Pi$--$\Delta\nu$ diagram. Similarly, the pulsational properties of some red giants do not match expectations based on their location in the HR~diagram assuming single-star evolution theory. \citet{Li+22NatAs} demonstrated that some red giants had higher masses and non-degenerate cores in the past, whereas today they have much lower masses, which can be explained by mass transfer with an as-yet-undetected companion. Therefore, ensemble forward asteroseismic modeling assuming single-star evolution tracks should be treated with caution in parts of the HR~diagram where binary interaction products are likely to be located.

\subsection{Compact pulsators}
\label{section: results: compact}

The compact pulsators are found in the late stages of stellar evolution, and include white dwarfs and subdwarfs. White dwarfs are the ultimate fate of all low- and intermediate-mass stars, and have a long history in the literature in terms of asteroseismology --- see \citet{WingetKepler08araa} for a review. Since white dwarfs are generally faint objects, binary studies can be challenging. For example, very few binaries containing two white dwarfs are known and those have only been discovered relatively recently (e.g.\ \citealt{Steinfadt+10apjl, Parsons+11apjl}).

\begin{marginnote}[]
\entry{Roche lobe overflow}{When a star overfills its Roche lobe and loses mass to its companion or from the system.}
\end{marginnote}

\begin{marginnote}[]
\entry{Common envelope}{A short-lived phase in the evolution of a binary star where the envelope of the more massive star engulfs the system as it expands on the way to becoming a supergiant.}
\end{marginnote}

Subdwarfs (sdO and sdB stars) have masses of about 0.5\Msun\ and are the products of binary evolution \citep{Heber09araa}. They are formed when a red giant is stripped of its envelope through a \emph{common envelope} ejection or stable \emph{Roche lobe overflow} mechanism \citep{Han+02mnras, Han+03mnras}. Subdwarfs were discovered to pulsate two decades ago and are usually either pressure- or gravity-mode dominated based on their effective temperatures \citep{Kilkenny07comms}. An example of a sdB binary is NY\,Vir, which has both primary and secondary eclipses, and high-amplitude multiperiodic pulsations in its discovery light curve \citep{Kilkenny+98mnras}.

A noteworthy example of a doubly compact binary system is KIC 7668647, which is an sdB star and a white dwarf in 14.2\,d orbit. Asteroseismology using space-based photometry combined with spectroscopy constrained interior rotation \citep{Telting+14aa}.





\section{IMPACT OF BINARITY ON PULSATIONS AND STELLAR STRUCTURE}
\label{section: impact}

The pulsations of stars in wide binaries with negligible tides are typically modeled using single-star evolution models. For dEBs, fundamental parameters from binary modeling, specifically masses and radii, are highly advantageous constraints on the subsequent forward asteroseismic modeling. However, the use of single star evolution theory for close binaries may not be wise, because tides can change a star's structure and evolution. Here we discuss how the structure and pulsations of binary stars can differ from single stars.

\subsection{Convection, mixing, and core masses}
\label{section: impact: mixing}

An important and currently not fully calibrated aspect of theoretical stellar models for all stars is the mixing length theory (MLT) of convection \citep{Bohmvitense58za}. This parameter quantifies the radial distance over which a volume of gas convects before coming into equilibrium with its surroundings, and is effectively an efficiency length scale for convection. However, convection is an inherently 3D process. Regardless, this mixing length, denoted as $\alpha_{\rm MLT}$, in 1D stellar evolution models is expressed in terms of the pressure scale height and directly controls the size of convective envelopes in low-mass stars \citep{JoyceTayar23galax}. For example, a larger $\alpha_{\rm MLT}$ in a low-mass dwarf yields a higher \Teff\ and smaller radius, which propagates as an uncertainty throughout a star's evolution. 

The Sun is an unique calibrator of $\alpha_{\rm MLT}$, with values in the range of $1.8 < \alpha_{\rm MLT, \odot} < 2.0$ reported in the literature depending on the physical ingredients of the models \citep{JCD02rimp}. A notable example of constraining $\alpha_{\rm MLT}$ for a star other than the Sun is the binary system $\alpha$\,Cyg\,A and B which contains a pair of solar analogs. Constraining $\alpha_{\rm MLT}$ was only possible through a combination of ultra-precise binary modeling, asteroseismology and interferometric constraints yielding $\alpha_{\rm MLT,A}/\alpha_{\rm MLT, \odot} = 0.932$ and $\alpha_{\rm MLT,B}/\alpha_{\rm MLT, \odot}=1.095$ \citep{Joyce++18apj}, which is not possible for the vast majority of stars. Importantly, $\alpha_{\rm MLT}$ is a prescription describing a 3D dynamical instability in a 1D evolution code, and is based on scaling the Sun's convective efficiency to other stars. Hence our knowledge of how valid the application of MLT to massive stars or evolved low-mass stars remains limited (see \citealt{JoyceTayar23galax}), but pulsating binaries clearly provide a route forward for testing MLT.

\begin{marginnote}
    \entry{Convective boundary mixing (CBM)}{Dynamical processes at the interface of convective and radiative zones which contribute to mixing of chemical species.}
\end{marginnote}

Another important and currently uncalibrated aspect of interior mixing inside stars is \emph{convective boundary mixing} (CBM), which occurs at the interface of convective and radiative zones. Various physical mixing processes at this interface can occur and are collectively referred to as CBM, such as waves, plumes, eddies, and convective penetration \citep[see][for a review]{AndersPedersen23galex}. The most commonly discussed mechanism is how plumes can overshoot the convective core boundary as defined by the Schwarzschild criterion because they have momentum, thus travel a certain distance before dissipating and causing chemical mixing. The term overshooting is therefore sometimes used in the literature synonymously with CBM, despite it describing only one of multiple physical scenarios. In early-type dwarfs, CBM is particularly important because it directly controls the effective size of the hydrogen-burning convective core. With a larger amount of mixing, additional hydrogen fuel is supplied to the core for nuclear fusion, thus changing the star's structure and lengthening its main-sequence lifetime by upwards of 25\% (see e.g.\ \citealt{Bowman20faas}). Moreover, CBM affects the mass of the helium core at the end of the main sequence \citep{Johnston21aa}, hence also supernova chemical yields and compact remnant masses \citep{Temaj+24aa}. 

CBM was introduced in evolution models to provide an improved match to the color-magnitude diagrams of open clusters as well as consolidate the observed and predicted locations of the terminal-age main sequence within the HR~diagram (e.g.\ \citealt{Castro+14aa}). Pulsating dEBs are particularly useful laboratories for calibrating CBM because they contain two stars of known mass, radius and luminosity, and the same age and initial chemical composition. This significantly improves asteroseismic modeling by having independent fundamental parameters from binarity to better constrain CBM using pulsations. 

An early study demonstrating the importance of CBM by \citet{Andersen++90apj} included 13 dEBs and two open clusters. \citet{Ribas++00mn} extended this work to 45 dEBs and found that the CBM increases with stellar mass, although by how much is uncertain \citep{Claret07aa}. Later, \citet{ClaretTorres16aa,ClaretTorres18apj} used 37 dEBs to show how CBM increases for stars with masses above 1.1\Msun\ until reaching a plateau for masses between about 2 and 5\,\Msun. However, similar analyses have questioned whether sufficient precision in CBM has been achieved \citep{ConstantinoBaraffe18aa} or is even achievable \citep{Valle+16aa} without additional information such as pulsations (see also \citealt{Tkachenko+20aa}).

\begin{marginnote}
    \entry{Mass discrepancy}{The significant difference in the masses of a binary system inferred based on spectroscopy and evolutionary models.}
\end{marginnote}

Effectively all evolutionary studies of binary systems have demonstrated the need to include CBM, especially when reconciling dynamical masses with evolutionary or spectroscopic masses. For example, \citet{Tkachenko+24aa} demonstrated using a sample of SB2 systems across a wide range in mass that the \emph{mass discrepancy} in massive binary systems is likely related to underestimated convective core masses provided by stellar evolution models (see also \citealt{Johnston21aa}). The dynamical masses of EBs allowed robust testing of whether the mass discrepancy lies in determining spectroscopic masses (i.e.\ from the RV semi-amplitudes of an SB2) or evolutionary masses (i.e.\ based on location of the stars in the HR~diagram). This emphasizes the need to exploit double-lined dEBs and, if pulsations are also present, perform asteroseismic modeling \citep{Tkachenko+20aa, Tkachenko+24aa, Johnston21aa}.

\subsection{Circularization, synchronization, rotation, and angular momentum transport}
\label{section: impact: AM transport}

It is well established theoretically how tides lead to circularization and synchronization of close binary stars \citep{Zahn1975aa, Zahn77aa, Zahn89aa, ZahnBouchet89aa, TassoulTassoul90apj, TassoulTassoul92apj}. For stars with radiative envelopes, a tidal torque is caused by the \emph{dynamical tide}, which can excite waves near the convective core that travel towards the stellar surface and then dissipate and deposit angular momentum \citep{Zahn1975aa, GoldreichNicholson89apj}. Extensions to the asymptotic theory of \citet{Zahn1975aa} are the inclusion of rotation \citep{Mathis09aa}, non-adiabaticity \citep{Savonije++95mn, SavonijePapaloizou97mn}, and resonances between the frequency of the tidal torque and a star's pulsation frequencies \citep{WitteSavonije99aa, Fuller17mn}. Moreover, it is also now possible to directly calculate the tidal torque in numerical stellar evolution computations \citep{Sun++23apj}.  

\begin{marginnote}
    \entry{Dynamical tide}{Time-dependent tidal force causing deformation of a star from the gravity of a companion, which is strongest at periastron passage in an eccentric binary.}
\end{marginnote}

Early observational evidence of synchronization caused by tides in close binaries was based on spectroscopy of relatively small samples (e.g.\ \citealt{Levato74aa}). Yet, recent space photometry has shown how the time scales and efficiency of tides depend on the orbital and stellar parameters \citep{JustesenAlbrecht21apj, Zanazzi22apjl,IJspeert+24aa}. Turbulent dissipation of the equilibrium tide leads to circularization for binary systems with periods up to about 2\,d during the pre-main sequence for cool stars with convective envelopes \citep{Barker22apjl}. Circularization for hot stars with radiative envelopes is instead achieved through radiative dissipation \citep{JustesenAlbrecht21apj}, or through an interaction with internal waves (see Section~\ref{section: impact: TEOs}). From a sample of about 14,500 EBs observed by the TESS mission and spanning a spectral type range of O-F dwarfs, \citet{IJspeert+24aa} find a small but significant reduction in the fraction of circularized short-period EBs with pulsations compared to non-pulsating EBs in the same orbital period range. Furthermore, \citet{Beck+24aa} discussed how the equilibrium tide is responsible for the location of a pulsating red giant in the $e$--$\log\,P$ diagram. This suggests that pulsations have a role in both circularization and synchronization of binaries. However, current observations are unable to determine a single mechanism that explains all properties of close binary stars across the required parameter space.

Primarily because of the complications that tides introduce on stellar structure and pulsation calculations, the sample of well-characterized pulsating EBs remains relatively small (see e.g.\ \citealt{Lampens21galax}). Among massive stars, there are very few pulsators in binary systems that have undergone forward asteroseismic modeling. \citet{Burssens+23natas} reported a differential radial rotation profile for the single $\beta$~Cep pulsator HD\,192575, with $\Omega_{\rm core}/\Omega_{\rm env} = 1.49^{+0.56}_{-0.33}$ ($2\sigma$ confidence intervals). \citet{Briquet+07mn} also reported quasi-rigid rotation based on forward asteroseismic modeling of the $\beta$~Cep star in the binary system $\theta$\,Oph\,A. The small sample size combined with an incomplete theory for angular momentum transport in single stars \citep{Aerts++19araa} makes drawing conclusions for the angular momentum history of massive binary systems particularly difficult (see \citealt{Marchant++24araa}).

At intermediate masses, \citet{Li+20mnras2} studied 35 eclipsing and 45 pulsation-timing binaries containing $\gamma$~Dor pulsators, observed during the \textit{Kepler} mission. These binary systems had gravity-mode period spacing patterns which allowed their near-core rotation rate to be constrained. \citet{Li+20mnras2} conclude that tides are responsible for the observed synchronization of the systems with orbital periods less than 10\,d. However, three of these short-period systems showed sub-synchronous near-core rotation rates. They conjectured that tidally excited oscillations (TEOs; see Section~\ref{section: impact: TEOs}) may be responsible for angular momentum transport from the deep stellar interior to the orbit. Later, \citet{Fuller21mn} provided a theoretical framework to explain these observed sub-synchronous rotation rates with \emph{inverse tides}. More recently, \citet{FullerFelce24mn} studied a sample of sub-synchronous pulsating stars in triple systems. The combination of spin and orbital precession and a suitable orbital period of the tertiary star gives rise to the inner binary exhibiting a \emph{sub-synchronous Cassini state}. Orbital evolution models show how tidal dissipation through the interaction of the tidal frequency and inertial waves leads to a spin-orbit misalignment, as well as slower than expected rotation rates for the inner pair of stars when only considering the equilibrium tide \citep{FullerFelce24mn}.

\begin{marginnote}[]
\entry{Inverse tides}{Transfer of angular momentum by tidally excited oscillations (TEOs) from the deep interior of a star in a binary system into the orbit.}
\entry{Sub-synchronous Cassini state}{Orbital precession and spin–orbit precession in a triple system leads to the lowest energy state being spin-orbit misalignment and sub-synchronous rotation for the inner binary.}
\end{marginnote}

\subsection{Tidally excited oscillations}
\label{section: impact: TEOs}

Tides in binary systems can cause (time-dependent) deformation of stars, thus altering their structure and evolution. The equilibrium tide is responsible for the ellipsoidal effect, and the dynamical tide causes a time-variable distortion of stars in eccentric orbits which is strongest at periastron passage. The periodic tidal distortion of eccentric ellipsoidal variables earned them the nickname of `heartbeat stars', because their light curves can resemble an electrocardiogram. The dynamical tide in eccentric ellipsoidal binaries can (but not always) act as a driving mechanism for \emph{tidally excited oscillations} (TEOs), which have frequencies at exact harmonics of the orbital frequency \citep{Kumar++95apj}. 

\begin{marginnote}[]
\entry{Tidally excited oscillations (TEOs)}{Pulsations in eccentric binary systems with frequencies that correspond to a harmonic of the orbital frequency driven by the dynamical tide at periastron passage.}
\end{marginnote}

The breakthrough study of TEOs was of KOI-54 (HD\,187091), which is an eccentric binary with an orbital period of 41.8\,d studied using \textit{Kepler} data and shown to have a peak in brightness at periastron passage \citep{Welsh+11apjs}. KOI-54 also shows two gravity modes with frequencies corresponding to exactly 90 and 91 times the orbital frequency: these are TEOs following the theory of \citet{Kumar++95apj}. The sample of eccentric binaries with ellipsoidal modulation was soon expanded using \textit{Kepler} data: \citet{Thompson+12apj} reported 17 systems, and 172 exist within the catalog of EBs published by \citet{Kirk+16aj}. An excerpt of the \emph{Kepler} light curve of KIC 3749404 is shown in Fig.~\ref{fig: example LCs}. The orbital periods of eccentric ellipsoidal variables are up to 100\,d, and eccentricities are typically high with values between 0.2 and 0.9. The theory of \citet{Kumar++95apj} excellently predicts the light curves of such systems for a range of values of $\Porb$, $e$ and $i$.

\begin{marginnote}[]
\entry{Resonance locking}{Feedback loop between stellar and orbital evolution in an eccentric binary system, which leads to TEOs being an exact harmonic of the orbital frequency.}
\end{marginnote}

The dynamical tide is important for understanding the orbital evolution of close binaries. This is because a coupling exists between the evolution of the pulsation mode frequencies of a star and the evolution of the binary orbital period. If a similar change between these exists it can lead to tidal \emph{resonance locking} (see \citealt{FullerLai12mn, Burkart++14mn, Fuller17mn}). For TEOs in eccentric ellipsoidal variables, a relationship exists whereby changes to pulsation modes caused by stellar evolution take place commensurately with tidal dissipation shortening the orbital period of the binary. This is why the frequencies of TEOs are always exact harmonics of the orbital frequency \citep{Fuller+17mn}. Just as high eccentricity is not always critical to create a sizable ellipsoidal modulation effect in a binary system, high eccentricities are also not always required for tidal resonance locking. For example, \citet{Cheng+20apj} found only two of four eccentric ellipsoidal variables in their sample to show evidence of tidal resonance locking. 

Stars with TEOs are typically, but not exclusively, intermediate-mass stars because of two selection effects: (i) intermediate-mass stars are (far) more common than massive stars; and (ii) TEOs, which are gravity modes, are not observable in low-mass main-sequence stars because of their thick convective envelopes. Notable examples of eccentric ellipsoidal variables with TEOs include KIC 4544587 which has $\Porb = 2.189$\,d, $e = 0.28$, and apsidal motion \citep{Hambleton+13mn}, and KIC 8164262 which has $\Porb = 87.5$\,d and $e = 0.87$ \citep{Hambleton+18mn}. These examples demonstrate how the phenomenon of TEOs exists across a wide range in $\Porb$ and $e$. \textit{Kepler} light curves of these systems revealed self-excited $\delta$~Sct pulsation frequencies, as well as high-amplitude TEOs at harmonics of the orbital frequency. \citet{Fuller+17mn} used asteroseismology to conclude that smaller amounts of CBM provided better fits to the observed amplitude of the TEO of KIC 8164262. Finally, the high eccentricity and spin-orbit misalignment of KIC 8164262 suggests that such systems are hierarchical triples \citep{Anderson++17mn}. 

Other important examples include KIC 3230227 \citep{Guo++17apj} and KIC 4142768 \citep{Guo+19apj}, which are EBs with TEOs and self-excited pulsation modes. Asteroseismology of the pulsations provided important constraints on the near-core and envelope rotation rates, as well as the roles of tidal dissipation, circularization and synchronization for these systems. Moreover, since both KIC 3230227 and KIC 4142768 are EBs they are excellent laboratories for constraining tidal theory within binaries with independently determined masses and radii. For more evolved systems, \citet{Beck+14aa} discovered 18 eccentric ellipsoidal variables containing a red giant, which had orbital periods up to 440\,d and eccentricities between 0.2 and 0.76. At the high-mass end of the main-sequence, HD\,177863 \citep{WillemsAerts02aa} and $\iota$\,Ori \citep{Pablo+17mn} are notable examples of binaries with TEOs. More recently, \citet{Kolaczek+21aa} used TESS data to discover 20 new eccentric ellipsoidal variables with primary component masses above about 2\Msun, seven of which have TEOs. \citet{Kolaczek-Szymanski2023a} showed theoretically that large changes to the orbital parameters of binary systems hosting TEOs can take place on short timescales (e.g.\ thousands of years). This makes ensemble analysis of TEOs an interesting and complex subfield of asteroseismology --- see \citet{Guo21frontiers} for a review.

\subsection{Tidally perturbed pulsators}
\label{section: impact: tidally perturbed}

There have been several theoretical models of how the equilibrium tide impacts the pulsations of stars in circularized and synchronized close binary systems (see e.g.\ \citealt{Chandrasekhar63apj, ChandrasekharLebovitz63apj, Denis72aa, Saio81apj, ReyniersSmeyers03aa, ReyniersSmeyers03aa2, Fuller+20mn}). However, direct testing of these theories had to await time-series data from space photometry missions (e.g.\ \citealt{Balona18mn, Handler+20natas, Steindl++21aa}). Generally speaking, \emph{tidally perturbed pulsators} are close binary systems for which the equilibrium tide is sufficiently strong to affect a star's pulsations in some manner (see e.g.\ \citealt{Saio81apj, Smeyers05aspc}). The specific impact on a star's pulsations depends on the properties of the binary system and the relative strength of the equilibrium tide versus other important factors, such as centrifugal deformation and the Coriolis force. 


\begin{marginnote}[]
\entry{Tidally perturbed pulsators}{Close binaries for which the equilibrium tide significantly impacts a star's pulsations.}
\end{marginnote}

Early theoretical work focused on how the equilibrium tide affects the frequencies of pulsation modes in circularized and synchronized close binaries with uniform radial rotation \citep{Denis72aa}. For example, \citet{ReyniersSmeyers03aa} provide a generalized mathematical framework for tidal splitting of pulsation frequencies in systems that are not necessarily synchronized (i.e.\ also include a dynamical tide), for cases when the pulsation axis is the rotation axis or the tidal axis (i.e.\ the line of apsides). For close binaries where the equilibrium tide needs to be considered, assuming the pulsation axis is aligned with the tidal axis rather than the rotation axis means one can define a set of additional mode frequencies arising from a single spherical harmonic, which are split by the orbital frequency set by the tidally perturbed azimuthal order, $\tilde{m}$ \citep{Smeyers05aspc}. For example, a single non-radial pulsation mode with an angular degree, $\ell$, in the star's co-rotating frame can be tidally split into a multiplet with $2\ell + 1$ components separated by the orbital frequency, but split further into multiplets with $\ell + 1$ components in the frame co-rotating with the orbit \citep{Balona18mn}. This tidal splitting of pulsation frequencies according to their $\tilde{m}$ value gives rise to complex multiplet structures in the observer's inertial reference frame. 

\begin{marginnote}[]
\entry{Tidally trapped pulsations}{Pulsations in close binaries that are confined to part of the stellar interior because of distortion by the equilibrium tide.}
\end{marginnote}

However, \citet{ReyniersSmeyers03aa} note that their first-order tidal splitting formalism does not include the Coriolis or centrifugal forces. Recent work by \citet{Fuller+25ApJ} solved for tidally perturbed pulsations including tidal distortion, centrifugal distortion, and the Coriolis force. They showed that modes of tidally distorted stars are not aligned with the tidal axis, but instead become \emph{triaxial pulsation} modes. The dipole modes become aligned with one of the three principal axes of the star (i.e.\ the x, y, or z-axis). These modes produce somewhat different phase and amplitude modulation than modes aligned with the tidal axis, which has been verified by recent observations \citep{Zhang+24mnras}.

\begin{marginnote}
    \entry{Triaxial pulsation}{Pulsation modes with an axis aligned with a principal (x, y, or z) axis of a star in a close binary because of tides.}
\end{marginnote}

When a star's tidal distortion becomes large, pulsation modes of different angular degrees become coupled. \citet{Fuller+20mn} studied the impact of the equilibrium tide on pulsation mode eigenfunctions in such highly distorted close binaries. If the equilibrium tide in a circularized and synchronized binary is strong enough to distort the shape of a star from being spherical to a Roche model geometry, this leads to pulsation mode eigenfunctions that differ from their spherical symmetry counterparts. \citet{Fuller+20mn} devised a new formalism to study the tidally distorted shapes of close binaries and demonstrated how coupling between modes of different $\ell$ values in an aspherical star can lead to pulsation modes being trapped within the tidal bulge. A superposition of pulsation modes of different $\ell$ with a pulsation axis aligned with the tidal axis can lead to the scenario of \emph{tidally trapped pulsations} --- for example, the pulsations have visible amplitudes only near the L1 or L3 Lagrange point of the pulsating star in a close binary \citep{Fuller+20mn}. 

The consequence of tidal trapping is periodic modulation of a pulsation mode's observed amplitude\footnote{Close binary systems with this amplitude modulation caused by tidal trapping have also been referred to as tidally tilted pulsators or single-sided pulsators in the literature, with a single pulsation frequency appearing as a multiplet separated by the orbital frequency because of the changing viewing angle of the pulsation during each orbit (see e.g.\ \citealt{Handler+20natas, Kurtz+20mn}). However, we emphasize that tidal splitting and tidal trapping both require a tilted pulsation axis.}, as well as periodic phase changes commensurate with the orbital period. This scenario is analogous to the oblique pulsator model devised for the magnetic roAp stars \citep{Kurtz90araa}, which show multiplets of pulsation frequencies split by the star's rotation frequency. Under certain conditions, the mode amplitudes in tidally trapped pulsators may be amplified because they can propagate much closer to the photosphere due to the lower gravity present at the L1 Lagrange point \citep{Fuller+20mn}. However, the development of a complete theory that includes coupling among pulsation modes of different angular degrees, as well as the Coriolis and centrifugal forces, is complex and still to be developed.

\subsubsection{Tidally perturbed pressure-mode pulsators}
\label{section: impact: tidally perturbed pressure}

Soon after the launch of the \textit{Kepler} mission, tides were recognized to play an important role in interpreting pulsating binary systems (e.g.\ \citealt{Hambleton+13mn, Guo+16apj}). For example, \citet{Balona18mn} recognized that the multiplets split by the orbital frequency in KIC 4142768 might be explained by the theory of tidally perturbed pulsation modes of \citet{ReyniersSmeyers03aa, ReyniersSmeyers03aa2}. However, the \textit{Kepler} sample of close binaries was limited and it was only since the launch of TESS that an expansive data set for probing how tides impact pulsations has become available. 

\begin{figure}
    \centering
    \includegraphics[width=\textwidth]{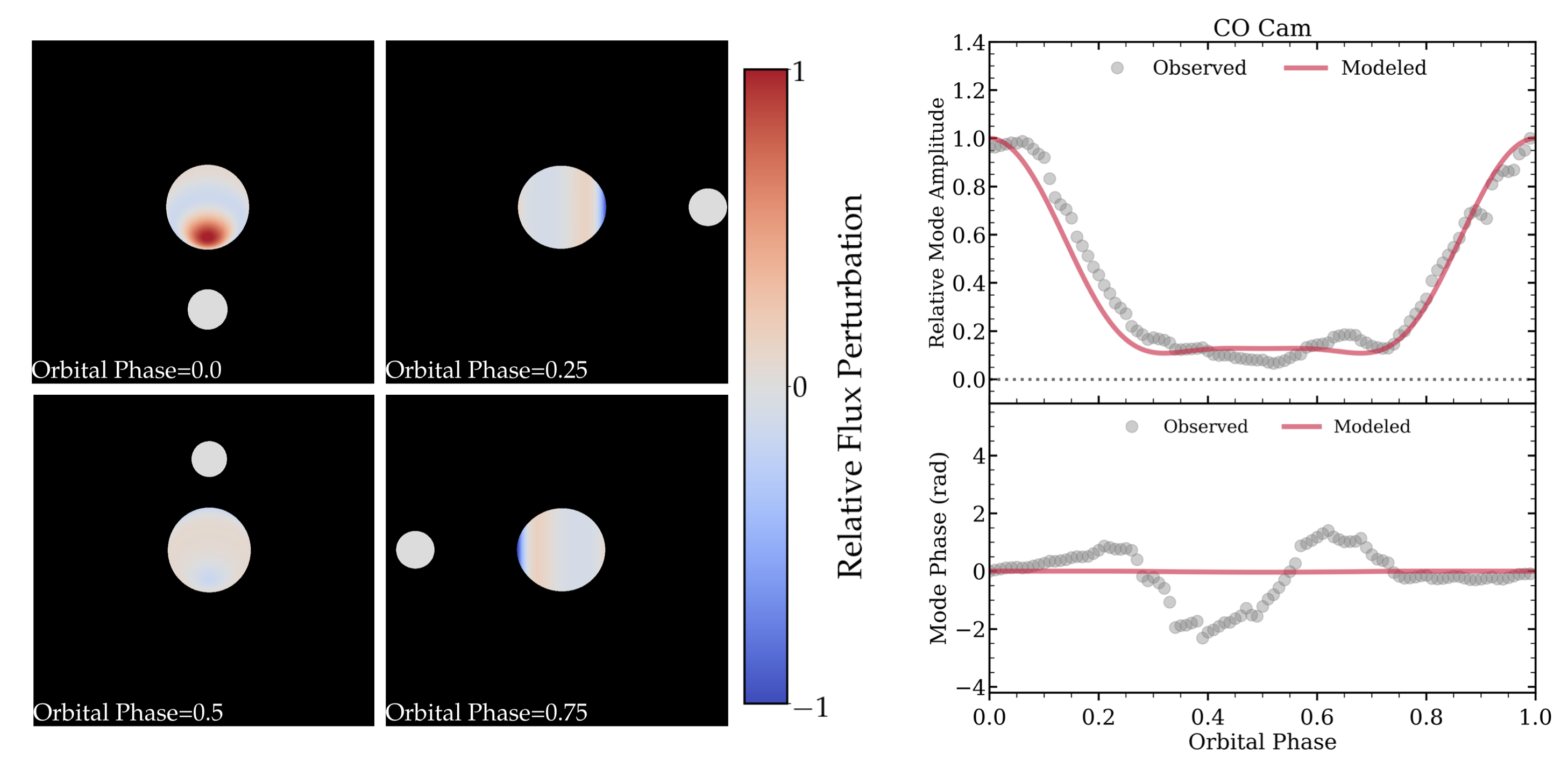}
    \caption{Left panels show 3D models of the tidally trapped pulsation at $\nu = 13.38$~d$^{-1}$ in CO~Cam \citep{Kurtz+20mn}. The system is viewed from an inclination of 49~deg and is fixed on the center of mass of the $\delta$~Sct primary. The right panel shows the observed and modeled amplitude and phase modulation. Figure adapted from \citet{Fuller+20mn}.}
    \label{fig:Fuller+2020}
\end{figure}

A notable example of a tidally trapped pulsator observed by the TESS mission is HD\,74423, which comprises two $\lambda$~Boo stars with similar primary and secondary masses of about 2\Msun\ and an orbital period of 1.58\,d \citep{Handler+20natas}. An excerpt of the TESS light curve of HD\,74423 and corresponding amplitude spectrum are shown in Figs.~\ref{fig: example LCs} and \ref{fig: example freq}, respectively. One of the components is a $\delta$~Sct pulsator and a multiplet of frequencies split exactly by the orbital frequency was found using the TESS light curve. \citet{Handler+20natas} concluded that a zonal pressure mode has a larger amplitude on the L1 side of the pulsating star. Another example is CO\,Cam, which is an Am-star primary and a G~secondary in an orbital period of 1.27\,d with four pressure modes, with a pulsation axis aligned with the tidal axis leading to amplitude modulation and multiplets split by the orbital frequency \citep{Kurtz+20mn}. A 3D model and the observed and modeled amplitude and phase modulation of CO\,Cam is shown in Fig.~\ref{fig:Fuller+2020}. A third example is TIC\,63328020, which comprises a $\delta$~Sct primary and a G~companion in a 1.11-d orbit \citep{Rappaport+21mn}. The general problem is complex since tidal trapping of pulsation modes does not always go hand-in-hand with tidal tilting \citep{Fuller+20mn, Fuller+25ApJ}. 

Since tides are a complex phenomenon for pulsating stars in close binaries, they are not yet fully understood. For example, it is not clear why only some of the pulsations in a close binary are affected by tides. However, this is likely related to the complex parameter space, both in terms of stellar and orbital parameters, needed to describe such systems. For example, a complex spectrum of pulsation frequencies, which include multiplets both exactly and quasi-equally spaced by the orbital frequency as well as independent modes was discovered for the oEA system U\,Gru by \citet{Bowman+19apj}. A similar conclusion of finding both perturbed and unperturbed pulsation modes was found for the 1.67-d binary RS\,Cha, which contains two equal-mass pulsating pre-main sequence $\delta$~Sct stars in a circularized and synchronized close binary \citep{Steindl++21aa}. The combination of photometric variability and spectroscopic mode identification through LPVs allowed \citet{Steindl++21aa} to apply the theory of \citet{ReyniersSmeyers03aa} and \citet{Smeyers05aspc} to RS\,Cha, and demonstrate that the frequencies within some of the observed multiplets arise from tidally perturbed modes (i.e.\ tidal splitting) based on their inferred $\tilde{m}$ values, rather than multiplets arising purely from a changing geometry from a single pulsation mode with a tilted pulsation axis. 

The multi-periodic pulsations of U\,Gru \citep{Bowman+19apj}, RS~Cha \citep{Steindl++21aa}, and the recently-discovered triaxial pulsator TIC~184743498 \citep{Zhang+24mnras} exhibit a mix of unperturbed and tidally perturbed pulsation modes, as well as amplification of some pulsation modes during the orbit. These pulsating binaries are crucial systems for understanding tides in close binaries. They demonstrate the diverse range of tidal behavior in close binaries, including how different pulsation modes can have different axes.



\subsubsection{Eclipse mapping}
\label{section: impact: eclipse mapping}

Pulsating stars in EBs are subject to geometry considerations where the visible amplitude of a non-radial pulsation mode may change during eclipse as different parts of the stellar disc are blocked by the companion. This effect is called \emph{eclipse mapping} and leads to a multiplet of frequencies equally spaced by the orbital frequency \citep{Reed++05apj,BiroNuspl11mn,Johnston+23aa}. 
This geometrical effect depends on the spherical harmonic geometry of a pulsation mode, and the stars' relative sizes and the eclipse duration. Therefore, multiplets split by the orbital frequency caused by eclipse mapping are important for close binaries.

\begin{marginnote}[]
\entry{Eclipse mapping}{Modulation of a pulsation mode's amplitude caused by the geometrical effect of an eclipse changing the surface area coverage of the pulsating companion in the observer's line of sight.}
\end{marginnote}

The pulsating oEA system U\,Gru was shown to exhibit a tidally perturbed pressure-mode that is appreciably affected by eclipse mapping, as well as several other independent pressure modes that seem unaffected by the equilibrium tide \citep{Bowman+19apj, Johnston+23aa}. Moreover, eclipse mapping of the tidally perturbed mode in U\,Gru creates a complex and asymmetric multiplet structure with component frequencies separated by the orbital frequency. Follow-up spectroscopy of U\,Gru also allowed \citet{Johnston+23aa} to establish it as a hierarchical triple system which further complicates the pulsation analysis. Yet eclipse mapping was only significant for one pulsation mode because of its specific spherical harmonic geometry \citep{Bowman+19aa, Johnston+23aa}.

\subsubsection{Tidally perturbed gravity-mode pulsators}
\label{section: impact: tidally perturbed gravity}

Recently a sample of tidally perturbed gravity-mode pulsators (i.e.\ $\gamma$~Dor and SPB stars) in close binaries has been discovered \citep{Jerzykiewicz+20mn, Vanreeth+22aa, Vanreeth+23aa}. This was surprising given that the equilibrium tide is expected to predominantly affect the near-surface layers of stars in close binaries. The discovery of tidally perturbed gravity modes, which are most sensitive near the convective cores of $\gamma$~Dor and SPB stars, is a challenge to explain theoretically (see \citealt{Fuller+25ApJ}).

\subsubsection{Discussion}
\label{section: impact: discussion}

Recent observational studies have demonstrated that pulsating stars in close binaries can exhibit a mixture of different tidal phenomena affecting their pulsation modes. Importantly, the detection of a multiplet of frequencies split by the orbital frequency for a close binary system is typically not sufficient to distinguish between the scenarios of: (i) tidal splitting of pulsation modes into multiplets (e.g.\ \citealt{Denis72aa, Balona18mn}); (ii) a single pulsation mode with amplitude modulation creating a multiplet structure because of the observer's changing viewing angle (e.g.\ \citealt{Fuller+20mn}); and (iii) eclipse mapping in the case of a pulsating EB (e.g.\ \citealt{Reed++05apj}). Additional constraints, for example LPVs, are extremely useful \citep{Steindl++21aa, Johnston+23aa}.

We therefore propose the convention that tidally perturbed pulsators are a general category of close binary stars for which the equilibrium tide impacts the pulsation frequencies in a measurable way. This includes any amount of tilting of the pulsation axis, and/or tidal splitting of pulsation frequencies (e.g.\ \citealt{Smeyers05aspc, Balona18mn}). Subsets within this group include: tidally trapped pulsators (e.g.\ \citealt{Fuller+20mn}) which describe those with amplitude modulation and phase modulation with respect to the observer because a pulsation mode's surface amplitude depends on a star's Roche potential; and triaxial pulsators (e.g. \citealt{Fuller+25ApJ}) which have multiple pulsation axes. Regardless of the tidal scenario, the diverse group of tidally perturbed pulsators show great promise for asteroseismology. This is because pulsation mode identification and model-independent masses and radii are obtainable (see \citealt{Steindl++21aa, Rappaport+21mn}).

\subsection{Binary-interaction products and mergers}
\label{section: impact: mergers}

The dynamical masses and radii of EBs are valuable constraints for evolution modeling. However, these measurements may not directly constrain the evolutionary history of a binary system, especially if the components have lost or gained mass due to their binarity. The stars in such systems are termed \emph{binary interaction products}. Under certain conditions, binary stars can even merge leaving behind a seemingly single star: a \emph{merger product}. It is thus difficult to disentangle single stars from merger products, which is increasingly important for more massive stars because their multiplicity and the probability of binary interaction are larger \citep{Sana+12sci, Offner+23aspc, Marchant++24araa}.

The incidence of pulsations in A and F dwarfs in close binaries is relatively high (e.g. \citealt{LiakosNiarchos17mn, Murphy+18mn}), but this is not the case for red giant stars in close binaries. Pulsations in red giants appear to be easily suppressed by proximity to a companion \citep{Gaulme+16apj, Beck+18mnras}, which is potentially caused by such stars having increased magnetic activity. Post-mass transfer binary systems including RR~Lyr and Cepheid pulsators (e.g.\ \citealt{Pietrzynski+12Nat, Pilecki+17apj}) have also been discovered, indicating that a binary evolution channel is needed for evolved pulsators \citep{GautschySaio17mnras}. Moreover, the relatively new group of pulsators known as blue large-amplitude pulsators (BLAPs), which are found on the blue side of the main sequence in the HR~diagram, have pulsation properties arising from being binary interaction products \citep{Pietrukowicz+17natast, ByrneJeffery20mnras, Byrne++21mnras}. Finally, the sdO and sdB stars have long since been established as binary interaction products \citep{Han+02mnras, Han+03mnras}, but only a fraction of them have pulsations.

\begin{marginnote}[]
\entry{Binary interaction product}{The outcome after binary components interact, such that the stars cannot be treated using single star evolution theory.}
\entry{Merger product}{A star that has resulted from the merging of two former stars.}
\end{marginnote}

\subsubsection{Algols and oEA binaries}
The oEA stars are EBs containing a $\delta$~Sct star that has been rejuvenated by mass transfer from a close companion \citep{Mkrtichian+02aspc, Mkrtichian+03aspc}. Asteroseismology of such binary interaction products is able to constrain the efficiency of mass transfer and angular momentum transport \citep{Miszuda+21mn, Miszuda+22mn}. For example, KIC 8262223 is an EB with a period of 1.6\,d, and its \textit{Kepler} light curve shows ellipsoidal modulation in addition to pulsations in the $\delta$~Sct primary and a low-mass evolved secondary \citep{Guo+17apj}. The secondary star of KIC 8262223 was constrained to have a mass of 0.20\Msun\ and radius of 1.31\Rsun. Yet this was not always so, since the present-day secondary was formerly the more massive star in the binary, meaning that it evolved more quickly and eventually filled its Roche lobe. At that point, it transferred a sizable fraction of its mass to the present-day primary, effectively doubling its mass to the current value of about 1.9\Msun\ \citep{Guo+17apj}. Forward asteroseismic modeling of pulsations of the $\delta$~Sct primary in KIC 8262223 using binary evolution models allowed the efficiency of the previous epoch of mass transfer to be constrained. The pulsation frequencies of KIC 8262223 were appreciably higher than expected for its mass and radius compared to other $\delta$~Sct stars. This suggests that the mass transfer event of helium-rich material occurred relatively recently and rejuvenated the star to become a (field) blue straggler \citep{Guo+17apj}. 

Another well-studied pulsating binary interaction product is KIC 7385478 \citep{Guo+19apjl}, which has a $\gamma$~Dor primary with gravity modes rather than the pressure modes typical of the higher-mass $\delta$~Sct stars. Since gravity modes directly probe the near-core of early-type dwarfs, \citet{Guo+19apjl} were able to show that the near-core rotation rate of the primary was synchronized with the orbit. Furthermore, in the synchronized EB KIC 9592855, the secondary star pulsates in gravity and pressure modes, which allowed a quasi-rigid radial rotation profile to be measured \citep{Guo++17apj2}. These examples are likely on the EL CVn evolution pathway, which includes a stage when binaries have an F or A dwarf as primary and a low-mass helium white dwarf precursor as a companion \citep{Maxted+13mn, Maxted+14mn}. This binary formation channel is based on the evolution of two low-mass stars with stable Roche lobe overflow and a \emph{mass ratio reversal}. Therefore, after donating a large fraction of its initial mass, the original primary star can becomes an extremely low-mass white dwarf precursor pulsator, with a mass of about 0.2\Msun. 

\begin{marginnote}[]
\entry{Mass ratio reversal}{When mass transfer between stars in a binary system causes a reversal in labels for each star as primary and secondary.}
\end{marginnote}

\subsubsection{Be stars}

Be stars make up around 20\% of all B dwarfs and are defined by having emission lines in spectroscopy. They are rapid rotators with transient decretion disks of ejected material \citep{Rivinius++13aapr}. The incidence of pulsations in Be stars correlates with increasing rotation rate, meaning Be stars are typically observed with non-radial gravito-inertial modes \citep{Labadie-Bartz+22aj}. The combination of rapid rotation and pulsations is thought to create an efficient angular momentum transport mechanism, but also provide the necessary dynamical mechanism required to eject material from the star into their decretion disks \citep{Huat+09aa, Kurtz+15mn, Neiner+20aa}.

However, the evolutionary pathway to form a Be star remains under debate. There is a lack of main-sequence companions to Be stars earlier than spectral type B1.5 in short-to-moderate period binaries (i.e.\ $\Porb < 5000$\,d; \citealt{Bodensteiner++20aa}). This supports a scenario for Be stars as binary interaction products, having gained mass and angular momentum from a (former) companion. If the initially more-massive companion since exploded as a supernova and provided a sufficient dynamical kick to disrupt the Be star from its orbit, it may become ejected from its companion leaving a runaway star seen today as a single Be star with a high proper motion (see \citealt{Marchant++24araa}).

\subsubsection{Stellar mergers}
Under certain conditions, for example mass transfer with a companion, or dynamical interaction with a third body (e.g.\ \citealt{Toonen+20aa}), it is possible for stars to merge and leave behind a merger product. The incidence of mergers is around 20\% for massive stars \citep{deMink+13apj, Henneco+24aa}, which are commonly born in multiple systems (e.g.\ \citealt{Sana+12sci, Offner+23aspc, Marchant++24araa}). A merger scenario is a highly disruptive event in binary evolution and depends strongly on how much mass is lost from an interacting binary system. It can also lead to the internal shear layers and a large-scale magnetic field \citep{Schneider+19nat, Frost+24sci}. 

Since pulsations are sensitive to a star's rotation and chemical profile, asteroseismology can distinguish merger products from single stars. \citet{Deheuvels+22aa} and \citet{Li+22NatAs} demonstrated that the pulsations of some red giants in binary systems reveal previous evolutionary phases of interaction. Moreover, \citet{RuiFuller2021mnras} identified 24 candidate mergers among red giant stars. At higher masses, \citet{Wagg+24aa} and \citet{Miszuda++25arXiv} studied how mass transfer leaves a detectable imprint on pulsations, which can be used to constrain the evolutionary history of binary systems. Whereas, \citet{Bellinger+24apjl} and \citet{Henneco+24aa, Henneco+25aa} demonstrated how gravity modes can distinguish single stars from merger products among a variety of massive stars, from the main-sequence through to the blue supergiants (see \citealt{Bowman+19natas}). Although merger seismology is an emerging field and its complexity cannot be understated, it has great potential to improve our understanding of binary star evolution across the HR~diagram.


\section{CONCLUSIONS: SUCCESSES AND ONGOING ISSUES}
\label{section: conclusions}

Pulsating binaries are excellent laboratories for testing and improving a plethora of physical processes in stellar structure and evolution theory. Since dEBs can be treated effectively as single stars they allow us to constrain single star evolution theory, whereas close binaries provide insight into the physics of binary star evolution. In the past few decades, the field of pulsating binary systems, and in particular pulsating EBs, has greatly benefited from plentiful high-quality space photometry. This has facilitated the discovery and analysis of much larger samples of pulsating EBs across the HR~diagram than previously possible, as well as allowing the highest-precision and model-independent constraints on masses and radii of stars in EBs, which are better than 1\% in the best cases. When combined with forward asteroseismic modeling, the model-independent dynamical masses and radii of EBs are effective in breaking degeneracies among stellar structure and evolution models.

A consensus has emerged that pulsations and binarity often need to be dealt with congruently when analyzing space mission light curves, especially when pulsation amplitudes are of similar amplitude to the eclipse depths. High-precision space photometry has also revealed that tides play an important role in interpreting the pulsations of binary systems. Not only do tides perturb pulsation frequencies from their single-star counterparts, they contribute to the deformation of a star's structure, and can lead to additional rotation and mixing processes. Tides are especially important for close binaries, and a diverse range of tidal phenomena has recently been discovered for different types of pulsating binaries across the HR~diagram, including TEOs in eccentric ellipsoidal variables, as well as tidally perturbed pulsations and triaxial pulsations in circularized and synchronized close binaries. 

\begin{summary}[SUMMARY POINTS]
\begin{enumerate}
\item Binarity is commonplace in the Universe, especially for higher mass stars, and plays a pivotal role in shaping a star's structure and evolution.
\item Eclipsing binaries (EBs) are excellent laboratories for testing stellar theory, as they provide empirical (model-independent) masses and radii.
\item Recent large-scale surveys, both photometric and spectroscopic, have led to a golden age for studying pulsations in binary stars.
\item Pulsating binaries can constrain mass transfer and angular momentum transport, and facilitate tidal asteroseismology for improving binary stellar evolution theory.
\end{enumerate}
\end{summary}

\begin{issues}[ONGOING ISSUES]
\begin{enumerate}
\item Pulsating EBs are relatively rare among all stars, and the relationship between their orbital and pulsational properties remains largely unconstrained.
\item Pulsating binaries generally require human intervention with choices of how to analyze the pulsations or the binarity. Ideally, the inclusion of pulsations self-consistently within a binary modeling code would avoid this subjectivity.
\item Atmospheric modeling codes are generally 1D, so stars deformed by tides or rapid rotation are affected by gravity darkening and may have incorrect parameters.
\item Binary modeling codes are generally slow because of their complexity, and estimating their internal errors is challenging.
\item In the era of excellent data, observers should pay particular attention to providing constraints that are best suited for testing theoretical model predictions.
\end{enumerate}
\end{issues}


\section{FUTURE OUTLOOK}
\label{section: future}

The NASA TESS mission is currently in its second extended mission and has been hugely successful for the fields of exoplanets, asteroseismology and EBs. High-precision light curves with a cadence of 200\,s are thus available for almost all stars brighter than about $V<13$\,mag. However, the vast majority of stars have a limited number of TESS sectors, each of length about 27.5\,d, owing to the mission design. This makes it challenging to detect and fully characterize EBs with periods longer than approximately 25\,d. A fraction of TESS targets are observed continuously for much longer when they lie in the continuous viewing zones (CVZs) at the two ecliptic poles, for which TESS light curves span 1\,yr \citep{Ricker+15jatis}. Therefore, stars within the TESS CVZs have maximal frequency resolution for combined asteroseismic and binary modeling studies.

The 4-year-long light curves of the \textit{Kepler} main mission are still yielding new scientific results: \textit{Kepler} data are arguably still the best source of time-series data for some groups of pulsating EBs due to the high photometric precision and long duration of the observations. This is particularly true for $\gamma$~Dor stars as these are relatively common within the \textit{Kepler} field of view, and the frequency resolution of a single sector of TESS data is generally insufficient to resolve individual gravity-mode frequencies, nor disentangle them from orbital harmonics. Further work in fully mining the \textit{Kepler} mission data will undoubtedly prove fruitful in the context of pulsating EBs containing intermediate-mass dwarf and giant stars. 

Another space mission that continues to revolutionize stellar astrophysics in general is ESA's \textit{Gaia} mission \citep{Gaia16aa, Gaia18aa, Gaia21aa}. The fourth data release (DR4) of \textit{Gaia} is expected in 2026 and will deliver a deluge of RV measurements for over 1 billion stars in the Milky Way, as well as full astrometric solutions and a photometric variability catalog. Moreover, \textit{Gaia}'s DR4 will include astrophysical parameters derived from the onboard BP/RP spectroscopy, as well as epoch data for potential exoplanet transits and EBs. Thus, plenty of new (pulsating) EBs are expected. The challenge will be to develop new automated tools to efficiently process and analyze such a vast data set.

The golden age of space photometry will continue. With an expected launch date in 2026, ESA's PLATO mission \citep{Rauer+25} will target brighter stars than the \textit{Kepler} mission did, which are much easier for RV follow up. Moreover, the PLATO fields of view are observed for much longer than a single sector of TESS data. Therefore, the synergy between TESS and PLATO is fantastic, with PLATO providing long-duration light curves for a pre-selected field of view, and TESS providing, on average, shorter light curves for the whole sky. Moreover, PLATO's CCDs are sensitive to bluer wavelengths than TESS, so for stars that are observed contemporaneously with the two missions multi-color light curves may be possible. This is a significant advantage for EBs and helps in constraining flux-dependent parameters such as \Teff, gravity darkening and contaminating light. 

In addition to large space photometry missions, a full exploitation of archival and new ground-based photometric and spectroscopic data to extend the baseline of time series data should be undertaken. This is motivated by the need to maximize the frequency resolution to study pulsations but also to sample different orbital period regimes. Ground-based surveys useful for studying pulsating EBs include WASP, HAT, MASCARA, HATPI, Evryscope, the Argus array, Warwick's Digital Telescope, ASAS-SN, ATO, and ZTF. 
Moreover, small CubeSats such as the BRITE constellation \citep{Weiss+21uni} and the upcoming high-resolution optical spectroscopic time-series CubeSpec mission \citep{Bowman+22aa} offer cheaper alternatives to large and expensive international space missions, albeit usually at a reduced scope in terms of scientific goals and number of targets. The upside of having fewer targets is that a much more specialized approach, such as high-cadence monitoring, is possible. This means that although fewer stars are studied, they have rich datasets to exploit for various scientific aims.

\begin{issues}[SUMMARY OF UPCOMING OPPORTUNITIES]
\begin{enumerate}
\item Full exploitation of archival ground-based photometric and spectroscopic data to extend the baseline of time series data of EBs is yet to be performed.
\item \textit{Kepler} mission light curves are yet to be fully exploited for pulsating binaries.
\item The TESS mission will (hopefully) continue for many years, providing longer light curves, thus improving the synergy of binary and asteroseismology studies.
\item The ESA \textit{Gaia} mission's fourth data release (DR4) will deliver RV measurements for 1 billion stars, astrometric solutions, and a photometric variability catalog, allowing for potentially millions of EBs.
\item The ESA PLATO mission will be launched in late 2026 and target brighter stars than the \textit{Kepler} mission, thus allowing for easier RV follow-up of pulsating EBs.
\item Cubesats offer cheaper alternatives to larger and expensive space telescopes, and a high degree of flexibility for specialized observations of bright targets.
\end{enumerate}
\end{issues}

\section*{DISCLOSURE STATEMENT}
The authors are not aware of any affiliations, memberships, funding, or financial holdings that might be perceived as affecting the objectivity of this review.
\section*{ACKNOWLEDGMENTS}
The authors would like to thank Jim Fuller, Donald Kurtz, Gerald Handler, and Simon Murphy for useful discussions and feedback, {and offer their condolences to the family and friends of Patrick Gaulme, who spent much of his career working on pulsating binaries.} JS acknowledges support from STFC under grant number ST/Y002563/1. DMB gratefully acknowledges UK Research and Innovation (UKRI) in the form of a Frontier Research grant under the UK government’s ERC Horizon Europe funding guarantee (SYMPHONY; grant number: EP/Y031059/1), and a Royal Society University Research Fellowship (grant number: URF{\textbackslash}R1{\textbackslash}231631). The \textit{Kepler} and TESS data used in this paper were obtained from the Mikulski Archive for Space Telescopes (MAST) at the Space Telescope Science Institute (STScI) and are freely available via the MAST archive: \url{https://archive.stsci.edu/}. MAST is operated by the Association of Universities for Research in Astronomy, Inc., under NASA contract NAS5-26555. Support to MAST for these data is provided by the NASA Office of Space Science via grants NAG5-7584 and NNX09AF08G, and by other grants and contracts. Funding for the \textit{Kepler} mission was provided by NASA's Science Mission Directorate, and provided by the NASA Explorer Program for the TESS mission. 



\newcommand{\aapr}[1]{A\&ARv}
\newcommand{\aj}[1]{AJ}
\newcommand{\apj}[1]{ApJ}
\newcommand{\apjs}[1]{ApJS}
\newcommand{\apjl}[1]{ApJ Letters}
\newcommand{\mnras}[1]{MNRAS}
\newcommand{\nat}[1]{Nature}
\newcommand{\aap}[1]{A\&A}
\newcommand{\araa}[1]{ARA\&A}
\newcommand{\pasj}[1]{PASJ}

\bibliographystyle{ar-style2}
\bibliography{jkt,more}

\end{document}